\documentclass[pra,twocolumn,showpacs]{revtex4}
\usepackage{graphicx}
\usepackage{color}
\begin{document}

\title{Excitation spectrum of a mixture of two Bose gases confined in a ring potential with interaction asymmetry}

\author{A. Roussou$^1$, J. Smyrnakis$^2$, M. Magiropoulos$^2$, Nikolaos K. Efremidis$^1$, and G. M. Kavoulakis$^2$}
\affiliation{$^1$Department of Applied Mathematics, University of Crete, GR-71004, Heraklion, Greece \\
$^2$Technological Education Institute of Crete, P.O. Box 1939, GR-71004, Heraklion, Greece}
\author{P. Sandin$^3$, M. \"Ogren$^3$, and M. Gulliksson$^3$} 
\affiliation{$^3$School of Science and Technology, \"Orebro University, 70182 \"Orebro, Sweden} 
\date{\today}

\begin{abstract}

We study the rotational properties of a two-component Bose-Einstein condensed gas of distinguishable atoms which
are confined in a ring potential using both the mean-field approximation, as well as the method of diagonalization
of the many-body Hamiltonian. We demonstrate that the angular momentum may be given to the system either via 
single-particle, or ``collective" excitation. Furthermore, despite the complexity of this problem, under rather 
typical conditions the dispersion relation takes a remarkably simple and regular form. Finally, we argue that under 
certain conditions the dispersion relation is determined via collective excitation. The corresponding many-body 
state, which, in addition to the interaction energy minimizes also the kinetic energy, is dictated by elementary
number theory.

\end{abstract}
 \maketitle

\section{Introduction}

In the heart of superfluidity -- which includes a whole collection of phenomena -- are  
non-classical rotational properties and the support of persistent currents \cite{SF}. One of the easiest 
and idealized models for the study of these properties is the one where the particles move in a ring 
potential. This model focuses on the longitudinal degrees of freedom and assumes periodic boundary 
conditions.

Remarkably, several recent experiments on Bose-Einstein condensed gases of atoms have managed to realize,
at least when the transverse degrees of freedom do not play a crucial role \cite{Patrik}, such a system 
and to investigate their rotational properties. More specifically, experimentalists have managed to trap 
atoms in toroidal/annular traps and have even created persistent currents in them \cite{Kurn, Olson, 
Phillips1, Foot, GK, Moulder, Ryu, WVK}. 

Going one step further, the addition of an extra, distinguishable, component may also be considered. This 
problem is even more interesting. The extra degrees of freedom associated with this extra component introduces 
novel and highly non-trivial effects. Interestingly enough, this has also become possible experimentally 
\cite{Zoran}. 

On the theoretical side \cite{2comp, 2compp0, 2compp1, 2compp2, 2compp3, 2compp4, 2compp5, 2compp6, 2compp7}, 
the general problem of a Bose-Einstein condensate with two distinguishable components -- which we label as 
$``A"$ and $``B"$ -- that is confined in a ring potential may be attacked at various levels of 
complication/difficulty. Assuming equal masses $M$ for the two components, there are two cases that one 
may distinguish. The first is the ``symmetric" one, where the scattering lengths $a_{AA}$, $a_{BB}$, and 
$a_{AB}$ for elastic atom-atom collisions between $AA$, $BB$, and $AB$ atoms respectively are all equal 
to each other. The second (and more realistic) is the ``asymmetric" one, where at least two of the scattering 
lengths are not equal to each other.

In the symmetric case the dispersion relation is exactly linear within the mean-field approximation \cite{2comp}
for $0 \le L \le N_B$ and $N_A \le L \le N = N_A + N_B$, where $L \hbar$ is the total angular momentum of the system, 
and $N_A$, $N_B$ are the numbers of particles in each component (here we assume without loss of generality that $N_B 
< N_A$). In the asymmetric case, the linearity of the spectrum disappears \cite{2compp6, 2compp7}, while for $N_B \le 
L \le N_A$ in both the symmetric and the asymmetric case the dispersion relation is more complex. 

In the present study we focus on the asymmetric case and use both the mean-field approximation, as well as the
method of diagonalization of the many-body Hamiltonian to study the rotational properties of this system. Two 
crucial assumptions are made throughout the paper. The first is that the inter- and intra-component effective 
interaction is repulsive. The second is that the two components coexist spatially. The condition for phase 
coexistence in a finite ring has been derived in Ref.\,\cite{2comp} and we make sure we do not violate it with 
any set of parameters that we use. Roughly speaking this condition demands that the repulsion within the same 
species is stronger than that of the different ones. 

In a real experiment there is also of course the question of dynamic instability. As shown in 
Ref.\,\cite{2comp} the condition of energetic stability coincides with the one of dynamic stability of the system. 
We should also mention that in spinor condensates realistic dynamic simulations show that the spatial separation 
of the two components is possible, and this may affect significantly the rotational behaviour of the system, see, 
e.g., Ref.\,\cite{2compp2}.

According to the results which are described below, under rather typical conditions, the minority component carries 
the majority of the angular momentum in the whole interval $0 < L \le N_B$. One of the novel results of our study 
is that under certain conditions the whole excitation spectrum is quasi-periodic (in addition to the periodicity 
dictated by the Bloch theorem \cite{FB}, which holds also in a two-component system \cite{2comp}) and may be 
derived from the one for $0 < L \le N_B$ by exciting the center of mass motion, either of the $A$ component, of 
the $B$ component, or both. 

Furthermore, in the limit of ``strong" interactions there is a very simple candidate state that minimizes the 
interaction energy of the system (under the assumption that there is no phase separation). This is the one where 
the density is homogeneous in each component separately, i.e., the one where the two order parameters $(\Psi_A, 
\Psi_B)$ of the two components are in the plane-wave states $(\phi_m, \phi_n)$. Here $\phi_m(\theta) = 
e^{i m \theta}/\sqrt{2 \pi R}$ are the eigenstates of the non-interacting problem, where $R$ is the radius of 
the ring, which have an angular momentum $m \hbar$. The corresponding total angular momentum in the pair of 
states $(\phi_m, \phi_n)$ is $L \hbar$, with $L = m N_A + n N_B$. A suitable choice of the integers $m$ and $n$ 
allows us to give any value to $L$, provided that $N_A$ and $N_B$ are relatively prime. Clearly, among all the 
possible values of $m$ and $n$ that satisfy the constraint of the angular momentum, one has to choose the pair
of $(\phi_m, \phi_n)$ that minimize the kinetic energy. 

The number-theoretic arguments presented above hold for any atom number. For large atom numbers the mean-field 
approximation provides an excellent description of the state of the system. Still, within the mean-field 
approximation one fixes the population imbalance $x_i = N_i/N$, treating $x_i$ as a continuous variable. Even 
though the number-theoretic behaviour that results from the analysis presented above still applies, it has more 
dramatic effects in the limit of small atom numbers. To explore these finite-$N$ effects, we use the method of 
numerical diagonalization of the many-body Hamiltonian. 

In what follows below we describe in Sec.\,II the model that we use and the two approaches, namely the mean-field
approximation and the diagonalization of the many-body Hamiltonian. In Sec.\,III we study the excitation spectrum, 
starting with the limit of long-wavelength excitations. In the same section we then focus on the mean-field 
approximation and show how one can derive the excitation spectrum starting from the one for $0 \le L \le N_B$. 
Then, in Sec.\,IV we investigate the excitation spectrum beyond the mean-field approximation, diagonalizing the 
many-body Hamiltonian. We first present an alternative way of exciting the system collectively and present an 
approximate generalization of Bloch's theorem. Finally, we compare the results that we get from the diagonalization 
with the ones of the mean-field approximation. In Sec.\,V we present a conjecture about the form of the many-body 
state that is expected to be the state of lowest energy under some conditions that are analysed. Finally in 
Sec.\,VI we give a summary of our study and an overview of our results.

\section{Model and approach}

The Hamiltonian that we consider is 
\begin{eqnarray}
   {\hat H} = \frac {\hbar^2} {2 M R^2} \sum_m m^2 ({\hat c}_m^{\dagger} {\hat c}_m
+ {\hat d}_m^{\dagger} {\hat d}_m)  
\nonumber \\
+ \frac 1 2 g_{AA} \sum_{i,j,k,l} {\hat c}_i^{\dagger} {\hat c}_j^{\dagger} {\hat c}_k {\hat c}_l
\, \delta_{i+j, k+l}  
\nonumber \\
+ \frac 1 2 g_{BB} \sum_{i,j,k,l} {\hat d}_i^{\dagger} {\hat d}_j^{\dagger} {\hat d}_k {\hat d}_l
\, \delta_{i+j, k+l} 
\nonumber \\
+ g_{AB} \sum_{i,j,k,l} {\hat c}_i^{\dagger} {\hat d}_j^{\dagger} {\hat c}_k {\hat d}_l
\, \delta_{i+j, k+l} , 
\end{eqnarray}
Here $\hbar^2 m^2/(2 M R^2)$ is the eigenenergy of the single-particle eigenstates $\phi_m(\theta)$. The mass $M$ is 
assumed to be the same for the two species, while $g_{ij} = U_{ij}/(2 \pi)$, with $U_{ij}$ being the matrix elements 
for zero-energy elastic collisions between the $AA$, $BB$, and $AB$ components. Also, ${\hat c}_m$ and ${\hat d}_m$ 
are the operators which destroy an $A$, or a $B$ atom with angular momentum $m \hbar$, respectively. In what follows 
below we set $E_m = m^2 \epsilon$, where $\epsilon = \hbar^2/(2 M R^2)$ and also $\hbar = 2M = R = 1$. 

In the case of a single component this problem has been attacked by Lieb and Liniger \cite{LLM, LM} with 
use of the Bethe ansatz. The case of two species with equal scattering lengths has also been considered, see, e.g.,
Refs.\,\cite{BA}. In the present study we attack this problem in two ways. The first is within the mean-field 
approximation, introducing the two order parameters $\Psi_A$ and $\Psi_B$ of the two components, thus solving the 
corresponding coupled, Gross-Pitaevskii-like equations, (with $\Psi_A$ and $\Psi_B$ normalized to unity),
\begin{eqnarray}
  - \frac {\partial^2 \Psi_A} {\partial \theta^2} + (U_{AA} N_A |\Psi_A|^2 + U_{AB} N_B |\Psi_B|^2) 
  \Psi_A &=& \mu_A \Psi_A
  \nonumber \\
  - \frac {\partial^2 \Psi_B} {\partial \theta^2} + (U_{BB} N_B |\Psi_B|^2 + U_{AB} N_A |\Psi_A|^2) 
  \Psi_B &=& \mu_B \Psi_B,
\nonumber \\
\end{eqnarray}
where $\mu_A$ and $\mu_B$ is the chemical potential, and $N_A$ and $N_B$ is the number of atoms in each component. 
We find the solutions of lowest energy of the above equations imposing the constraint of some fixed angular momentum, 
as described in detail in Ref.\,\cite{2compp7}.

Alternatively we solve this problem by diagonalizing the many-body Hamiltonian. Within this scheme we choose a set 
of single-particle states $\phi_m(\theta)$, with $m_{\rm min} \le m \le m_{\rm max}$, making sure that a decent
convergence has been achieved with respect to $m_{\rm min}$ and $m_{\rm max}$. In this subspace of basis states 
we impose the constraints of a fixed number of atoms $A$ and $B$, $N_A$ and $N_B$, respectively. We also impose 
the constraint of some fixed angular momentum $L$ (which can be shared between the two components), see, e.g., 
\cite{2comp}. Finally we diagonalize the resulting Hamiltonian matrix in this subspace, thus deriving the 
eigenstates and the corresponding eigenenergies. 

The terminology of the ``yrast" state that we use below refers to the eigenstate with the lowest eigenenergy, i.e., 
the state which minimizes the energy for some given eigenvalue of the angular momentum. The same term is used within 
the mean-field approximation, where one fixes the expectation value of the angular momentum, instead. 
We stress that these yrast states play a fundamental role in the rotational response of these systems, very much
like the phonon-roton spectrum in the problem of liquid Helium. Finally, we also stress that the problem of fixing 
the angular momentum is intimately connected with the one where the angular velocity of the trap is fixed, instead. 

\section{Excitation spectrum -- Mean-field approximation}

\subsection{Elementary excitations}

Let us start with the mean-field approximation. When the system has zero angular momentum, $L=0$, it is in the state 
\begin{eqnarray}
|L = 0 \rangle = |0^{N_A} \rangle_A \bigotimes |0^{N_B} \rangle_B, 
\end{eqnarray}
where in this notation we have $N_A$ and $N_B$ atoms in the single-particle state with $m = 0$. The total energy 
of the system is 
\begin{eqnarray}
 E_0 = \frac 1 2 g_{AA} N_A (N_A - 1) + g_{AB} N_A N_B 
 \nonumber \\
 + \frac 1 2 g_{BB} N_B (N_B - 1).
\end{eqnarray}
Giving one unit of angular momentum via single-particle excitation to, e.g., the $B$ component, then 
\begin{eqnarray}
|L = 1 \rangle = |0^{N_A} \rangle_A \bigotimes |0^{N_B-1}, 1^1 \rangle_B,
\label{siexc}
\end{eqnarray}
and correspondingly for the species $A$. The total energy of this state is 
\begin{eqnarray}
 E' = 1 + \frac 1 2 g_{AA} N_A (N_A - 1) + g_{AB} N_A N_B 
 \nonumber \\
 + \frac 1 2 g_{BB} [(N_B-1) (N_B - 2) + 4 (N_B-1)].
\end{eqnarray}
Therefore,
\begin{eqnarray}
 E' - E_0 = 1 + g_{BB} (N_B - 1),
 \label{diffe}
\end{eqnarray}
where the last term comes from the exchange interaction. From the above equation it follows that it is the ratio 
\begin{eqnarray}
 r = \frac {g_{BB} (N_B-1)} {g_{AA} (N_A-1)}
 \label{ratio}
\end{eqnarray}
which determines whether the angular momentum goes to the one, or the other component. 

In what follows below we set $g_{AA} = g_{BB} = g$, and thus as Eq.\,(\ref{ratio}) implies, with the assumption 
$N_A > N_B$ that we have made, we conclude that it is the $B$ (minority) component that carries the angular momentum, 
for $L=1$. By the way, Eq.\,(\ref{diffe}) may be identified as the speed of sound $c$ of the $B$ component, or 
equivalently as the slope of the dispersion relation for $L \to 0^+$ for exciting it. More specifically,
\begin{eqnarray}
 c = 1 + g (N_B - 1).
 \label{diffee}
\end{eqnarray}  

\subsection{Distribution of the angular momentum between the two components}

While the above result holds for $L=1$, it turns out that more generally, under ``typical" conditions (which will be
analysed below) the minority component carries the largest part of the angular momentum, all the way up to $L=N_B$. 

The two order parameters may be expanded in the basis of plane-wave states,
\begin{eqnarray}
 \Psi_A = \sum_{m = m_{\rm min}}^{m = m_{\rm max}} c_m \phi_m, \,\,\, 
 \Psi_B = \sum_{m = m_{\rm min}}^{m = m_{\rm max}} d_m \phi_m.
\end{eqnarray} 
The corresponding energy per atom is
\begin{eqnarray}
  \frac E N = \sum_{m_{\rm min}}^{m_{\rm max}} m^2 (x_A c_m^2 + x_B d_m^2) 
   \nonumber \\ + \frac 1 2 x_A^2 N U \int \left| \sum_{m_{\rm min}}^{m_{\rm max}} c_m \phi_m \right|^4 \, d \theta
   \nonumber \\ + \frac 1 2 x_B^2 N U \int \left| \sum_{m_{\rm min}}^{m_{\rm max}} d_m \phi_m \right|^4 \, d \theta
   \nonumber \\ + x_A x_B N U_{AB} \int \left|\sum_{m_{\rm min}}^{m_{\rm max}} c_m \phi_m \right|^2 \, 
  \left| \sum_{m_{\rm min}}^{m_{\rm max}} d_m \phi_m \right|^2 \, d \theta.
\label{enen}
\end{eqnarray}

Considering the limit of weak interactions, in the interval $0 \le \ell \le 1$ one may work with the states with $m=0$ 
and $m=1$, only,
\begin{eqnarray}
  \Psi_A = c_0 \phi_0 + c_1 \phi_1, \,\,\, \Psi_B = d_0 \phi_0 + d_1 \phi_1,
\end{eqnarray}
where $\ell = L/N = x_A c_1^2 + x_B d_1^2$ is the angular momentum per particle. In the ``symmetric" case $(g = g_{AB})$ 
it turns out that for $0 \le \ell \le x_B$ \cite{2compp0},
 \begin{eqnarray}
  c_0^2 = \frac {(x_A - \ell) (1-\ell)} {x_A (1 - 2 \ell)}, \,\,\, c_1^2 = \frac {(x_B - \ell) \ell} {x_A (1 - 2 \ell)}
 \label{occup1}
 \end{eqnarray}
and 
\begin{eqnarray}
  d_0^2 = \frac {(x_B - \ell) (1-\ell)} {x_B (1 - 2 \ell)}, \,\,\, d_1^2 = \frac {(x_A - \ell) \ell} {x_B (1 - 2 \ell)},
\label{occup2}
\end{eqnarray}
with $c_0 c_1 d_0 d_1$ negative (as minimization of the energy implies). In this symmetric case the maximum value of the 
angular momentum carried by the majority component in the interval $0 \le \ell \le x_B$ is of order $x_B^2/4$, for $\ell 
\approx x_B/2$ (for relatively small $x_B$). Figure 1 shows $c_0^2, c_1^2, d_0^2$, and $d_1^2$ for $x_A = 0.8$ and $x_B 
= 0.2$. We have seen numerically that in the asymmetric model ($g > g_{AB}$) the angular momentum of the majority 
component decreases as $g/g_{AB}$ increases. This is expected, since in the limit of $g_{AB} \to 0$, the two components 
decouple. Thus, from the above expressions we can get an upper bound on the angular momentum carried by the majority 
component, which is $\approx x_B^2/4$, at least for reasonably small values of $x_B \le 0.3$. 

For stronger interactions (and still in the asymmetric case), as we have seen in our numerical results, the angular 
momentum carried by the majority component for $0 \le \ell \le x_B$ is still very small, on the order of 1\%, at least 
up to $x_B \le 0.3$ and $g/g_{AB} = 5/3$. 

Actually, we argue that this is a very general result, due to energetic reasons. There are four energy scales in the 
problem [see, e.g., Eq.\,(\ref{enen})], namely the kinetic energy (which is set equal to unity), the interaction energy 
among the $A$ particles, $\sim x_A^2 g N$, among the $B$ particles, $\sim x_B^2 g N$, and the interaction energy between 
the $A$ and the $B$ particles, $x_A x_B g_{AB} N$. There are thus three dimensionless parameters, namely the coupling 
$g$, the interaction asymmetry $g/g_{AB}$, and the population imbalance $x_A/x_B$. Clearly, for large values of 
$g/g_{AB}$ and/or large values of $x_A/x_B$, there is a clear hierarchy in the three energy scales of the interaction 
energy, which makes it energetically favorable for the system to carry its angular momentum by the one component (i.e., 
the $B$ component in this case).

\begin{figure}
\includegraphics[width=7cm,height=5cm,angle=0]{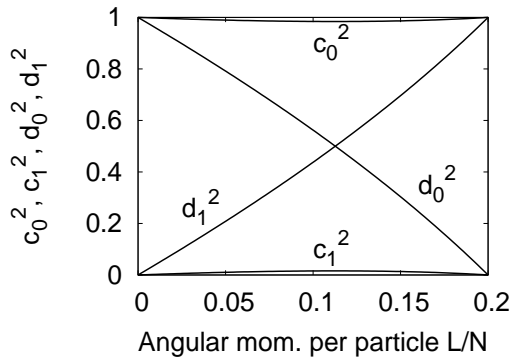}
\vskip3pc
\caption{The occupancy of the four states, $c_0^2, c_1^2, d_0^2$, and $d_1^2$, for $x_A = 0.8$ and $x_B = 0.2$ from
Eqs.\,(\ref{occup1}) and (\ref{occup2}). One can hardly distinguish the coefficients $c_0^2$ and $c_1^2$ from 1 and 0, 
respectively.}
\end{figure}

\subsection{Quasi-periodic structure of the dispersion relation and an explicit example with $x_A = 0.8, x_B = 0.2$}

As shown in Ref.\,\cite{2compp7}, for $x_A = 0.8$, $x_B = 0.2$, $N g/\epsilon = 1250/\pi^2$, and $N g_{AB}/\epsilon 
= 750/\pi^2$, to high accuracy the energy spectrum is given by the formula
\begin{eqnarray}
  E(\ell) = E_{\rm int} + P_0(\ell) + e_0(\ell).
  \label{fit1}
\end{eqnarray}
Here $E_{\rm int}$ is the interaction energy of the homogeneous system, $e_0(\ell)$ is a periodic function of $\ell$, and 
\begin{eqnarray}
 P_0 (\ell) = [\ell]^2 x_A + (\ell - x_A [\ell])^2/x_B,
\label{fit2}
\end{eqnarray}
where $[\ell]$ denotes the nearest-integer function. 

\begin{figure}
\includegraphics[width=7cm,height=5cm,angle=0]{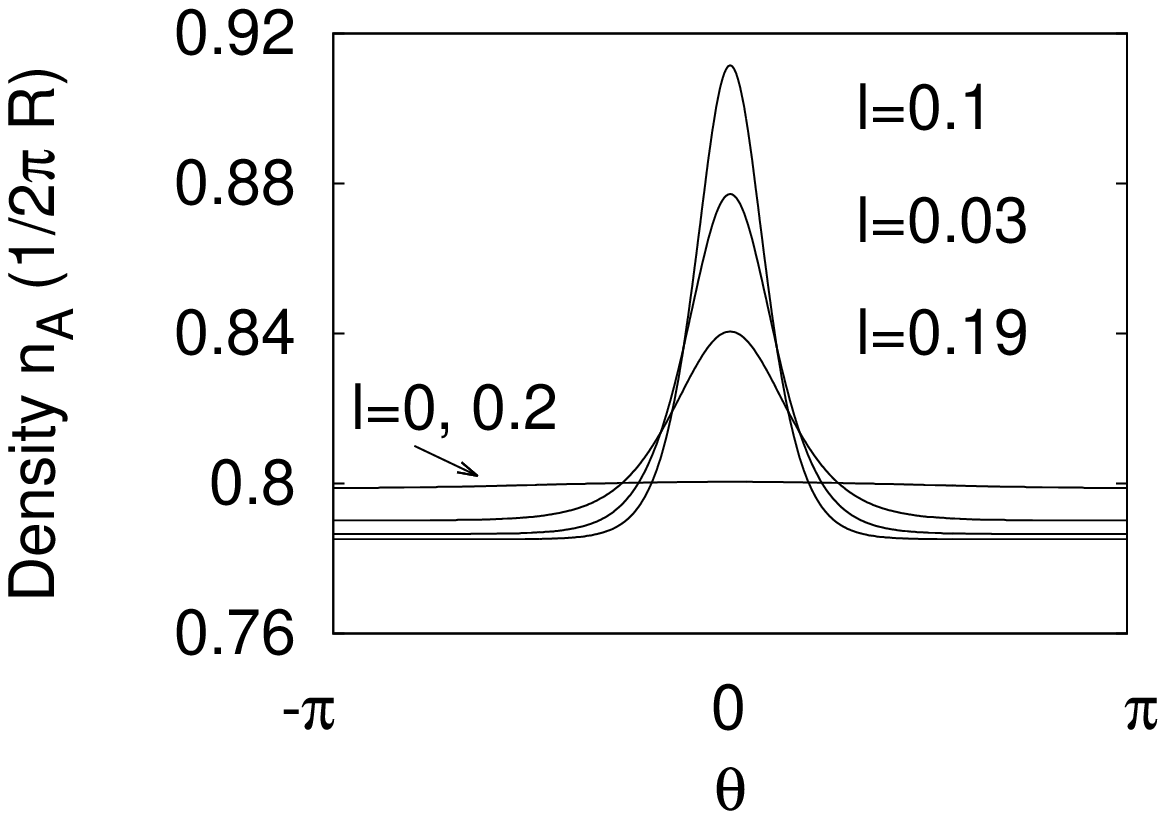}
\includegraphics[width=7cm,height=5cm,angle=0]{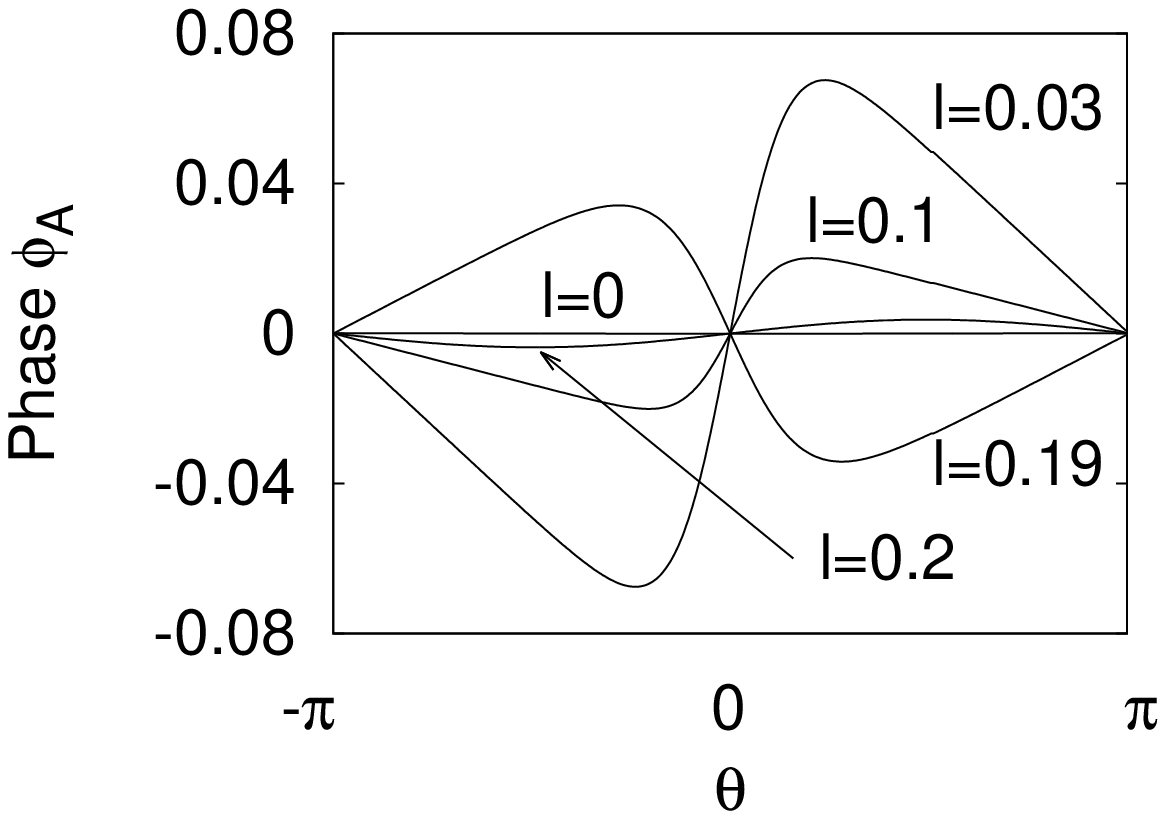}
\includegraphics[width=7cm,height=5cm,angle=0]{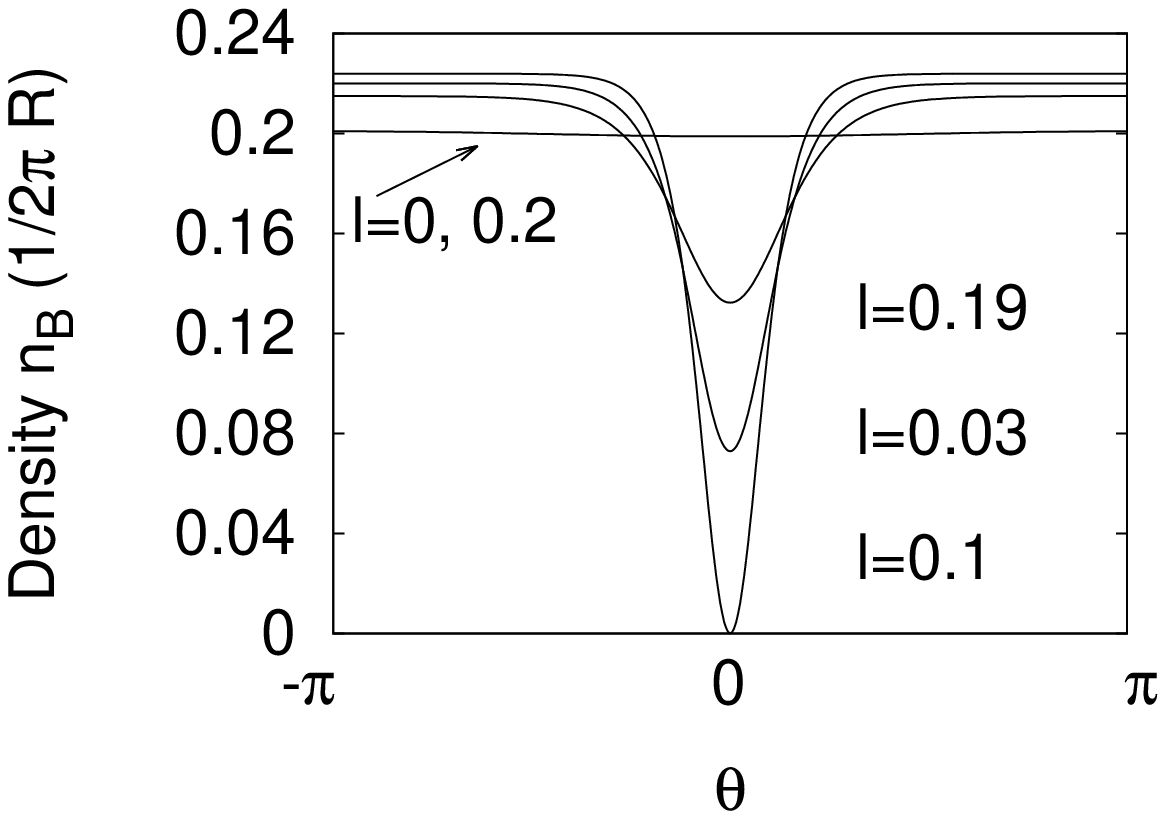}
\includegraphics[width=7cm,height=5cm,angle=0]{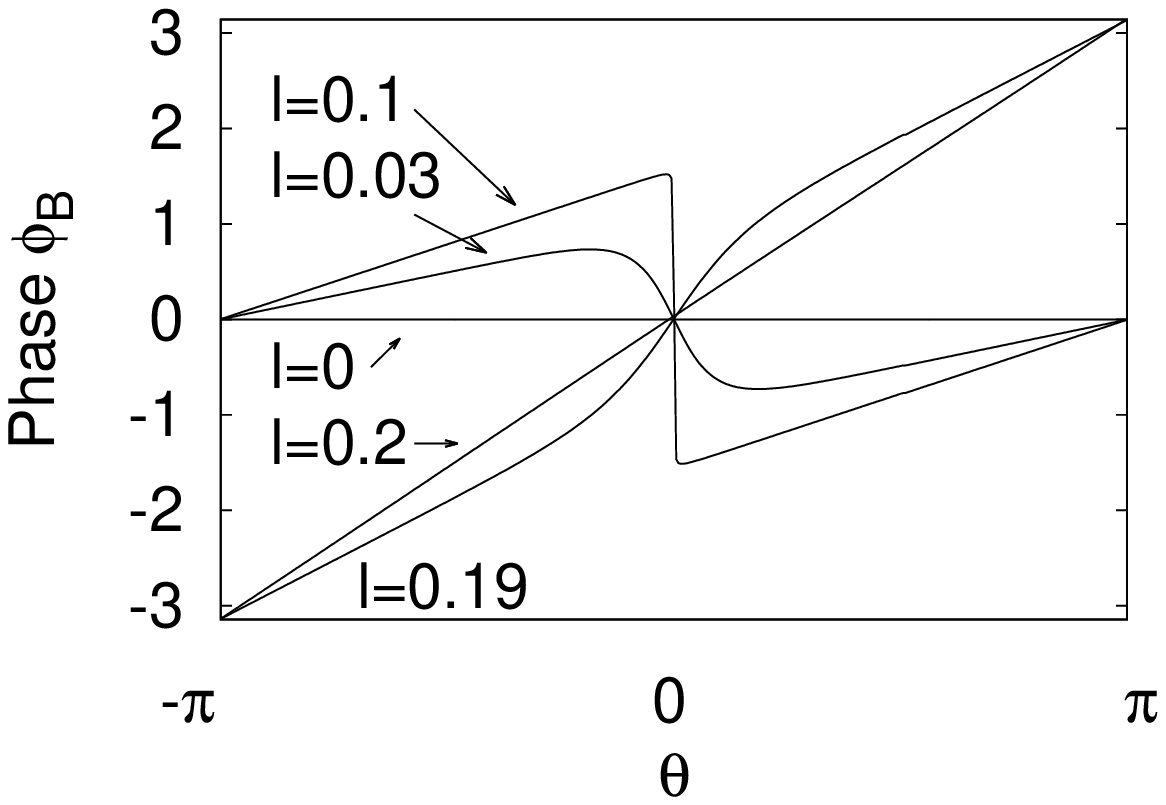}
\vskip3pc
\caption{Density and phase of the order parameters $\Psi_k = \sqrt{n_k} \, e^{i \phi_k}$ of the two components $A$ 
and $B$, for $x_A = 0.8$, $x_B = 0.2$ and for $\ell = 0, 0.03, 0.1, 0.19$, and 0.2. Here $N g/\epsilon = 1250/\pi^2$ 
and $N g_{AB}/\epsilon = 750/\pi^2$.}
\end{figure}

\begin{figure}
\includegraphics[width=7cm,height=5cm,angle=0]{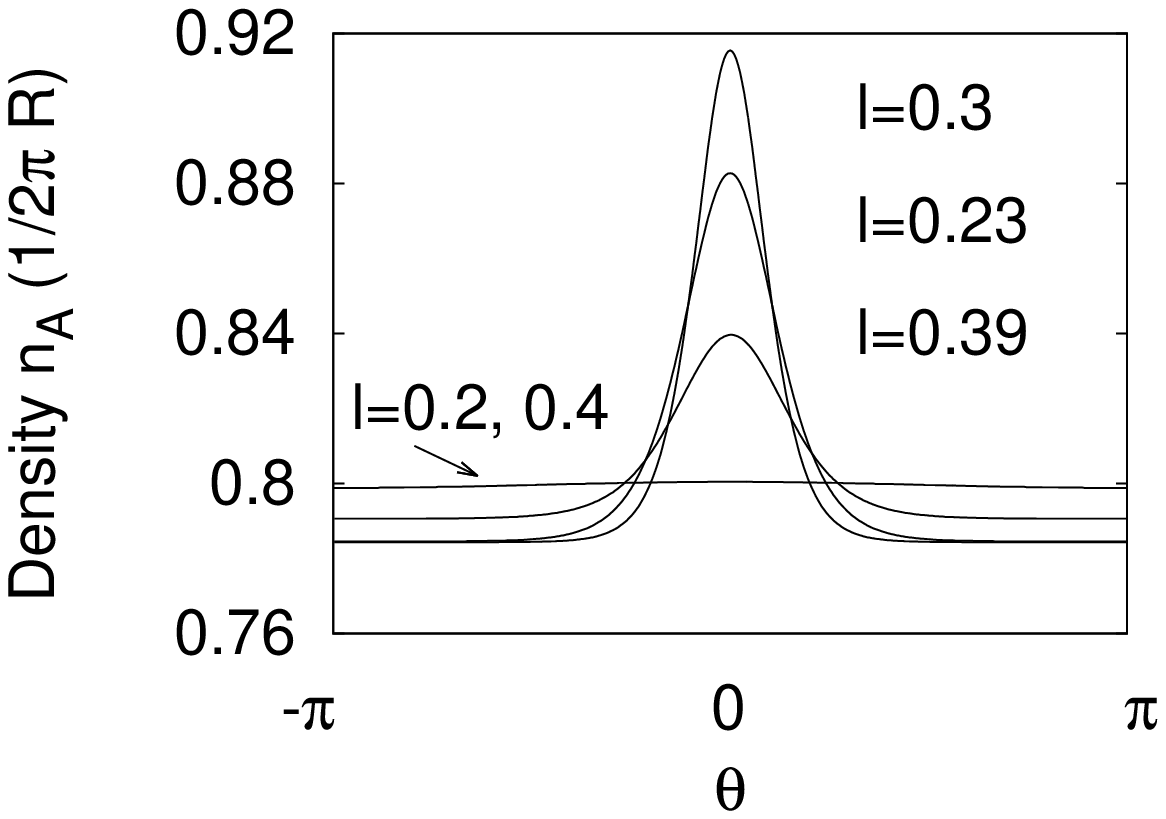}
\includegraphics[width=7cm,height=5cm,angle=0]{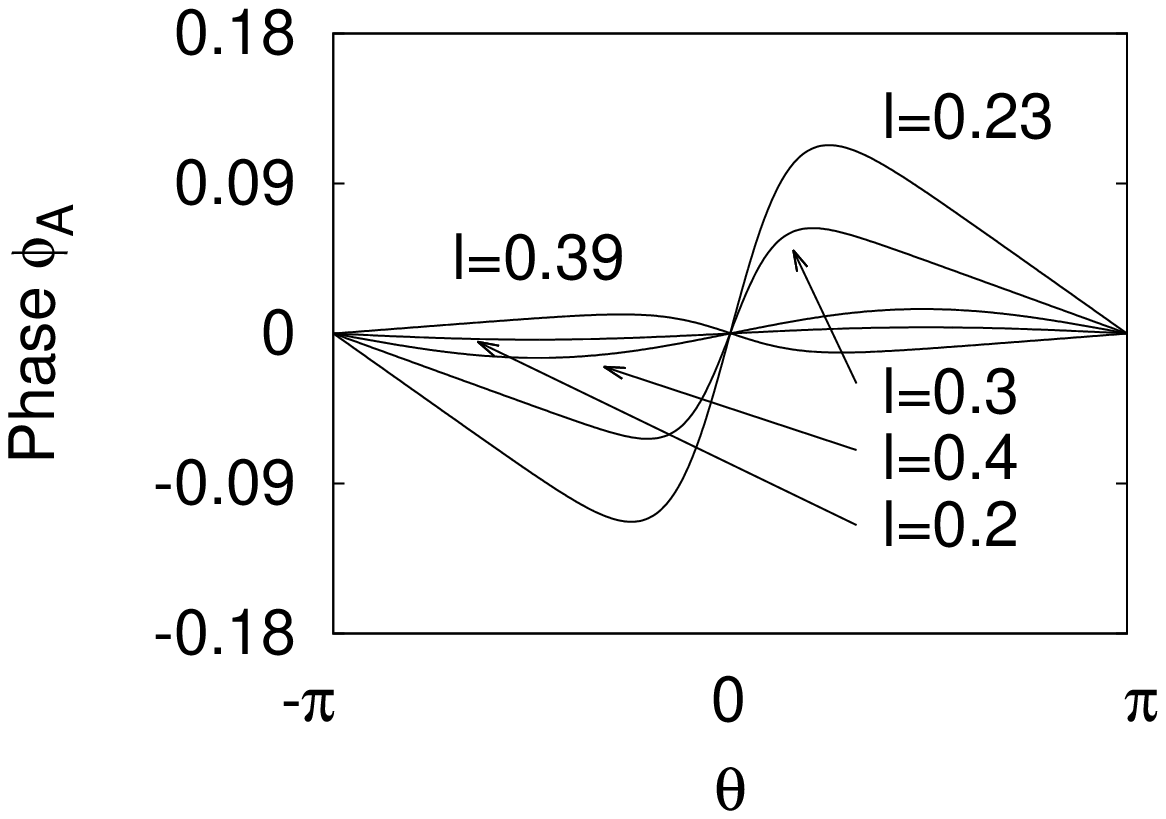}
\includegraphics[width=7cm,height=5cm,angle=0]{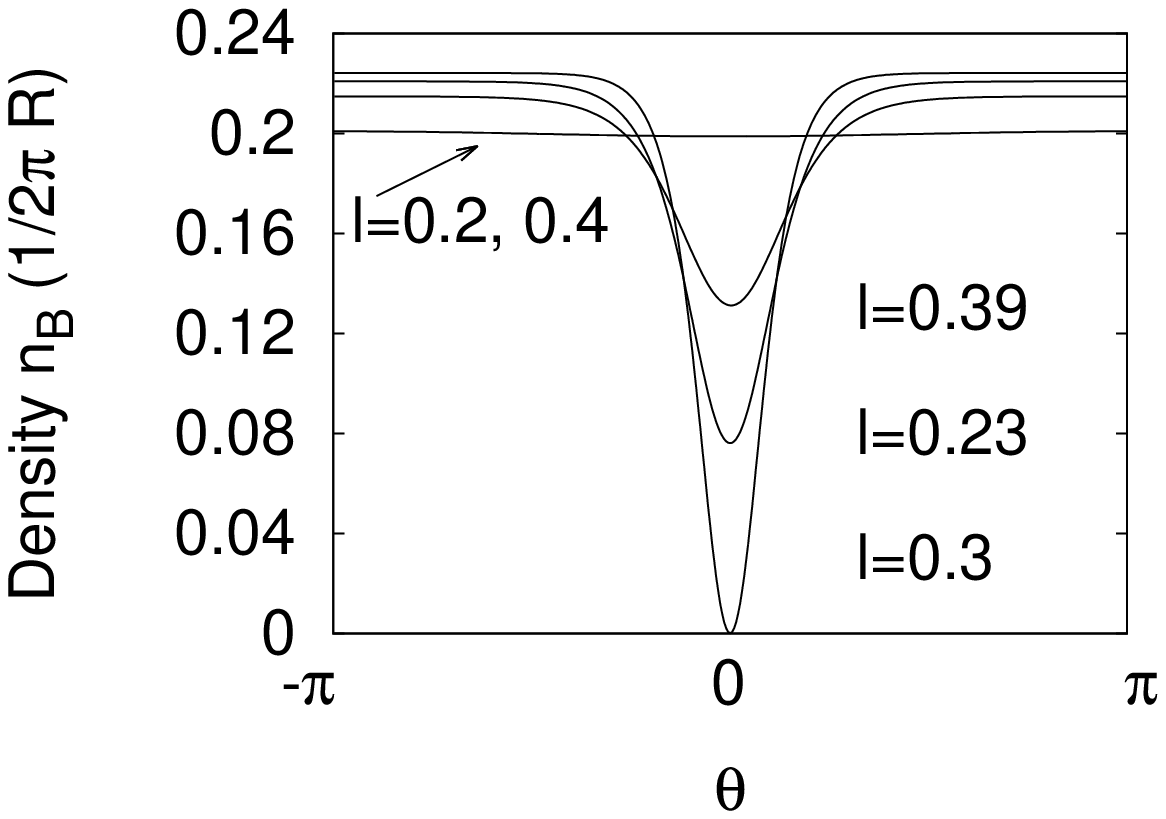}
\includegraphics[width=7cm,height=5cm,angle=0]{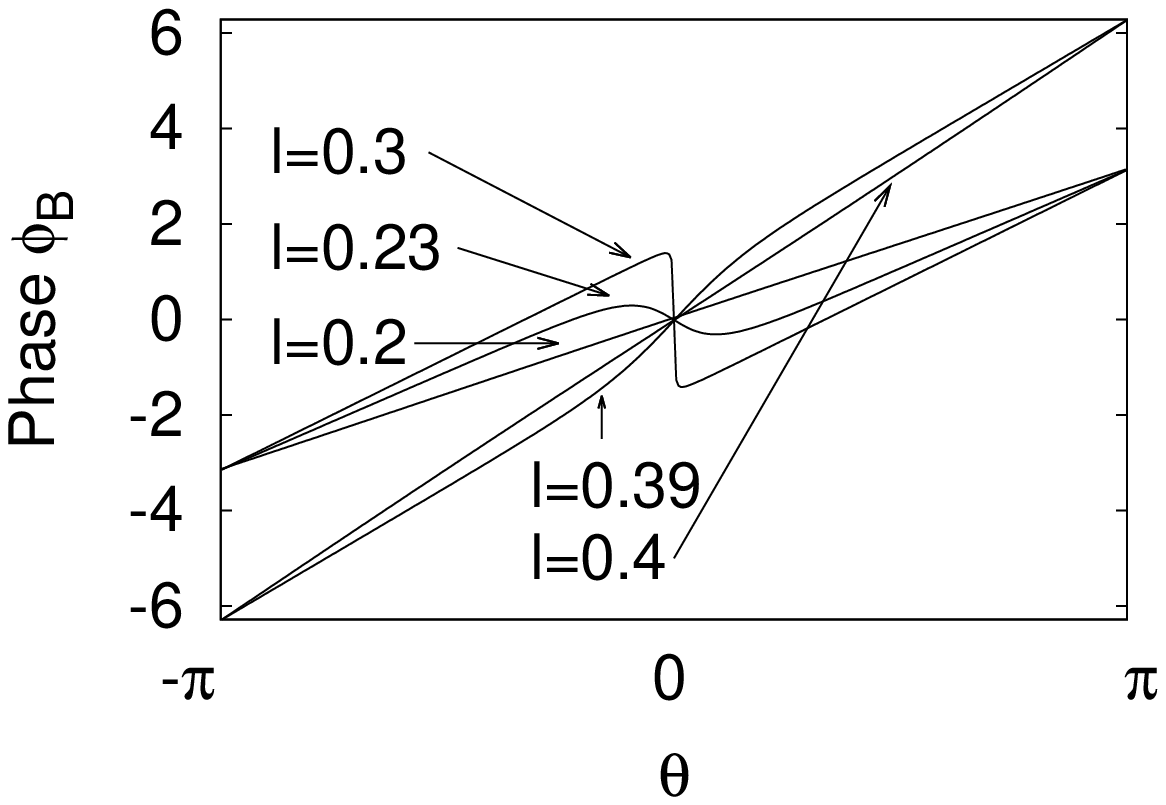}
\vskip3pc
\caption{Density and phase of the order parameters $\Psi_k = \sqrt{n_k} \, e^{i \phi_k}$ of the two components $A$ 
and $B$, for $x_A = 0.8$, $x_B = 0.2$ and for $\ell = 0.2, 0.23, 0.3, 0.39$, and 0.4. Here $N g/\epsilon = 1250/\pi^2$ 
and $N g_{AB}/\epsilon = 750/\pi^2$.}
\end{figure}

In Figs.\,2 and 3 we show the density and the phase of the two order parameters $\Psi_A$ and $\Psi_B$, in the two 
intervals $0 \le \ell \le 0.2$ and $0.2 \le \ell \le 0.4$. Comparing the density of the same species for values of 
$\ell$ which differ by $x_B = 0.2$ we observe that the difference is hardly visible. On the other hand, the phases 
of the two order parameters do change. These observations are explained in the analysis that follows below. Finally, 
the angular momentum carried by the majority component in the interval $0 \le \ell \le 0.2$ is very small, smaller 
than 1\%, as we argued also above. 
 
The above results follow from the facts that (i) at the interval $0 \le \ell \le x_B$ the minority component carries 
essentially all the angular momentum, and (ii) if one starts from the order parameters in the interval $0 \le \ell \le 
x_B$, the rest of the spectrum results by exciting the center of mass motion of each component separately. This 
operation changes the kinetic energy only, leaving the interaction energy unaffected. We thus essentially show below 
that Eqs.\,(\ref{fit1}) and (\ref{fit2}) follow from these two facts.

In order for the above procedure to give the yrast states, for a fixed population imbalance and a fixed interaction 
asymmetry, $g$ has to be sufficiently large. Considering, for example, $\ell = 0.4$, the yrast state -- which has to 
be $(\Psi_A, \Psi_B) = (\phi_0, \phi_2)$, as the quasi-periodic behaviour implies -- is indeed the expected one for a 
sufficiently strong interaction, as analysed in Ref.\,\cite{2compp6}. For a fixed interaction asymmetry and a fixed 
$g$, the population imbalance has to be sufficiently large. Finally, for a fixed $g$ and a fixed population imbalance 
the interaction asymmetry has to be sufficiently large. 

To see the above arguments it is instructive to consider the specific example $x_A = 0.8$, $x_B = 0.2$. 
First of all, the possible values of the angular momentum carried by (purely) plane-wave states is a multiple 
of 0.2 in this case, since $\ell = m x_A + n x_B = 0.2 (4 m + n)$. It is also important to notice that the 
condition for a state $(\Psi_m, \Psi_n)$ to become an yrast state, depends only on $|m-n|$ \cite{2compp6}. Thus, 
when, e.g., the state $(\Psi_m, \Psi_n) = (\phi_0, \phi_2)$ with $\ell = 2 x_B = 0.4$, becomes the yrast state, 
also the state $(\Psi_m, \Psi_n) = (\phi_1, \phi_{-1})$ with $\ell = 3 x_B = 0.6$, becomes the yrast state, as 
well. (This also follows from Bloch's theorem, however it is a more general result).

Having solved the yrast problem in the interval $0 \le \ell \le x_B = 0.2$, one may construct solutions at the
interval $0.2 = x_B \le \ell \le 2 x_B = 0.4$, etc., all the way up to $4 x_B \le \ell \le 5 x_B = 1$ keeping the
correlations unaffected and putting all the energy in the form of kinetic energy, by exciting the center of mass
motion. In other words, the spectrum will ``repeat" itself in a quasi-periodic way (explained below) all the way 
up to $\ell = 1$. Beyond this point Bloch's theorem determines the rest of the excitation spectrum \cite{2comp}. 

Let us thus assume that in the interval $0 \le \ell \le x_B = 0.2$ the two order parameters are 
\begin{eqnarray}
  (\Psi_A, \Psi_B) = (\Psi_A^0, \Psi_B^0). 
\end{eqnarray}
We should keep in mind that $\Psi_A^0$ carries a very small amount of angular momentum, and we will assume that 
it is zero. The angular momentum per particle of the above pair of states is $\ell = x_A \sum m c_m^2 + x_B \sum m 
d_m^2 = x_B \sum m d_m^2$, the kinetic energy per particle is $K^0(\ell) = x_A \sum m^2 c_m^2 + x_B \sum m^2 d_m^2$, 
and the total energy per particle is $E(\ell)/N = K^0(\ell) + V(\ell)/N$, where $V(\ell)$ is the total interaction 
energy. Finally, for the kinetic energy $K^0(\ell = 0) = 0$ and $K^0(\ell = x_B) = x_B = 0.2$.

For $0.2 = x_B \le \ell \le 2 x_B = 0.4$ the order parameters are
\begin{eqnarray}
  (\Psi_A, \Psi_B) = (\Psi_A^0, e^{i \theta} \Psi_B^0).
\end{eqnarray}
The factor that multiplies $\Psi_B^0$ does not affect the interaction energy and thus the interaction energy is
identical to the one in the interval $0 \le \ell \le x_B$, $V(\ell) = V(\ell - x_B)$. The interesting part is the 
kinetic energy, which is 
\begin{eqnarray}
K(\ell) = K^0(\ell-x_B) + 2 \ell - x_B,
\label{K1}
\end{eqnarray}
with $K(\ell = x_B) = x_B = 0.2$ and $K(\ell = 2 x_B) = 4 x_B = 0.8$. 

For $0.4 = 2 x_B \le \ell \le 3 x_B = 0.6$ we have two competing solutions around $\ell = 1/2$. For values of $\ell$ 
smaller than $1/2$, 
\begin{eqnarray}
  (\Psi_A, \Psi_B) =  (\Psi_A^0, e^{2 i \theta} \Psi_B^0).
\end{eqnarray}
The kinetic energy is 
\begin{eqnarray}
K(\ell) = K^0(\ell - 2 x_B) + 4 \ell - 4 x_B, 
\label{K2}
\end{eqnarray}
with $K(\ell = 2 x_B) = 4 x_B = 0.8$ and $K(\ell = 3 x_B) = 9 x_B = 1.8$. 

For larger values than $\ell = 1/2$,
\begin{eqnarray}
  (\Psi_A, \Psi_B) = (e^{i \theta} \Psi_A^0, e^{- 2 i \theta} \Psi_B^0).
\end{eqnarray}
The kinetic energy is 
\begin{eqnarray}
K(\ell) = K^0(\ell - x_A + 2 x_B) + 5 - 4 \ell - 9 x_B, 
\label{K3}
\end{eqnarray}
with $K(\ell = 2 x_B) = 5 - 17 x_B = 1.6$ and $K(\ell = 3 x_B) = 5 - 20 x_B = 1$. Comparing the energies one sees 
that they cross at $\ell = 5 x_A/8 = 1/2$. This gives rise to a discontinuity in the derivative of the dispersion 
relation at $\ell = 1/2$. We have evaluated the slope to be $1/x_B$ as $\ell \to (1/2)^-$ and $(1 - 2 x_A)/x_B$ for 
$\ell \to (1/2)^+$, and therefore the difference between the right and the left slopes is $-2 x_A/x_B$. 

We stress that this discontinuous transition at $\ell = 1/2$ is also experimentally relevant, since the slope of the 
dispersion relation gives the velocity of propagation of the corresponding solitary waves. Interestingly, at this 
point the sign of the slope changes and thus the velocity of propagation also changes sign.

For $0.6 = 3 x_B \le \ell \le 4 x_B = 0.8$,
\begin{eqnarray}
  (\Psi_A, \Psi_B) = (e^{i \theta} \Psi_A^0, e^{-i \theta} \Psi_B^0).
\end{eqnarray}
The kinetic energy is 
\begin{eqnarray}
K(\ell) = K^0(\ell - x_A  + x_B) + 1 - 2 \ell + 2 x_A - 2 x_B, 
\label{K4}
\end{eqnarray}
with $K(\ell = 3 x_B) = 1$ and $K(\ell = 4 x_B) = 0.8$. 

Finally, for $0.8 = 4 x_B \le \ell \le 1$, 
\begin{eqnarray}
  (\Psi_A, \Psi_B) = (e^{i \theta} \Psi_A^0, \Psi_B^0).
\end{eqnarray}
The kinetic energy is 
\begin{eqnarray}
K = K^0(\ell - x_A) + x_A, 
\label{K5}
\end{eqnarray}
with $K(\ell = 4 x_B) = x_A = 0.8$ and $K(\ell = 1) = 1$. Figure 4 shows the result of this calculation for $x_A = 0.8$
and $x_B = 0.2$.

The results presented above imply Eqs.\.(\ref{fit1}) and (\ref{fit2}), which were motivated numerically \cite{2compp7}, 
as mentioned also earlier. They are also consistent with the numerical results of Figs.\,2 and 3. We also stress that, 
although the arguments were presented within the mean-field approximation, they do not rely in any way on the validity 
of the mean-field approximation, but rather they are much more general, as we also demonstrate in Sec.\,IV. As a final 
remark we mention that when $N_A$ and $N_B$ are relatively prime, e.g., $x_A = 0.7$ and $x_B = 0.3$, a similar picture 
develops.

\begin{figure}
\includegraphics[width=7cm,height=5cm,angle=0]{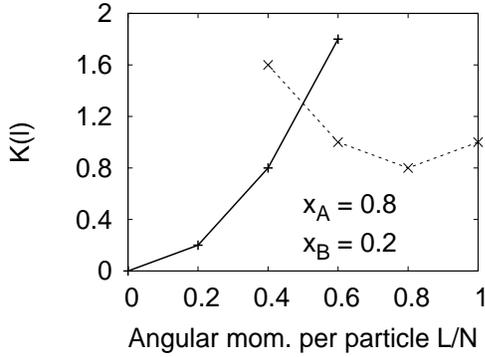}
\vskip3pc
\caption{The kinetic energy $K(\ell)$, evaluated at $\ell = 0, 0.2, 0.4, 0.6, 0.8$, and 1.0, for $x_A = 0.8$, 
and $x_B = 0.2$, from Eqs.\,(\ref{K1}), (\ref{K2}), (\ref{K3}), (\ref{K4}), and (\ref{K5}). Knowing the energy 
at the interval $0 \le \ell \le x_B$, one may derive the rest of the spectrum using the transformations described 
in the text.}
\end{figure}

\section{Excitation spectrum -- many-body problem}

\subsection{``Collective" excitation of the system}

Up to now we have seen how the yrast states progress with increasing angular momentum via essentially single-particle 
excitation of the system. In other words, as $L$ increases, the additional momentum is carried by moving single 
particles to different single-particle states. 

Still, there is another way to excite the system ``collectively". By this term we mean that even an increase 
of the angular momentum by one unit requires a major rearrangement of the atoms in the single-particle states. Before 
we go to the many-body problem, we should recall the results of Ref.\,\cite{2compp6}, where it was argued that for 
sufficiently strong interactions, the mean-field state $(\Psi_A, \Psi_B) = (\phi_m, \phi_n)$ becomes the yrast state, 
where obviously the angular momentum is $\ell = x_A m + x_B n$. 

A way to argue about the state $(\phi_m, \phi_n)$ becoming the yrast state for the specific value of $\ell$ and for 
sufficiently strong interactions is that any density variation costs interaction energy. If this is the dominant term 
in the Hamiltonian, it is minimized by these plane-wave states, which have a constant density distribution. The 
expense that one pays is the corresponding kinetic energy, which is $x_A m^2 + x_B n^2$, and has to be sufficiently 
small in order for the argument to be self-consistent; this argument is analysed further in Sec.\,V. The details of 
this calculation (performed within the mean-field approximation), as well as the corresponding phase diagram are 
given in Ref.\,\cite{2compp6}.

Let us thus consider a toy model which demonstrates the above arguments about the collective excitation. Assuming for 
convenience that $N_A - N_B = 1$, a state that competes with the one of Eq.\,(\ref{siexc}) is
\begin{eqnarray}
|L = 1 \rangle = |1^{N_A} \rangle_A \bigotimes |(-1)^{N_B} \rangle_B.
\label{coll} 
\end{eqnarray}
The energy of this state $E^{''}$ is 
\begin{eqnarray}
 E^{''} = N + \frac 1 2 g N_A (N_A - 1) + g_{AB} N_A N_B 
 \nonumber \\
 + \frac 1 2 g N_B (N_B - 1) = N + E_0,
\end{eqnarray}
or 
\begin{eqnarray}
  E^{''} - E_0 = N .
  \label{dief}
\end{eqnarray}
Therefore
\begin{eqnarray}
 E^{''} - E' = (N - 1) - g (N_B - 1).
\end{eqnarray}
For values of $g$ larger than the critical value which satisfies the equation
\begin{eqnarray}
 g = \frac {N-1} {N_B - 1},
\end{eqnarray}
it is energetically favourable to excite the system collectively. In the limit of large $N$ and $N_B$, $g$ is of 
order unity which is necessary in order for the system not to enter the highly-correlated Tonks-Girardeau regime.
(One should not forget that for the low atom numbers that we have used, the system easily makes the transition to 
the Tonks-Girardeau limit, when $g$ becomes of order $N$ \cite{TG}). 

We stress that the above calculation is just a toy model and should not in any way be trusted quantitatively. Besides,
for $g$ of order unity, the typical interaction energy per atom is of order $N$ and thus (much) larger than the 
kinetic energy. Thus, the interaction energy will deplete the condensate significantly, while the depletion will 
also make the result dependent on $g_{AB}$; all these effect have been ignored here.

\subsection{A ``generalization" of Bloch's theorem}

The arguments presented above ignore the depletion of the condensate. However, the depletion lowers the energy to 
subleading order in the number of atoms $N$ and, in particular for small systems, it may have a rather important
effect. Below, we suggest a different way of constructing a many-body state, taking into account also the depletion. 
Essentially this ansatz state generalizes (in an approximate way) Bloch's theorem, which also holds in a two-component 
system \cite{2comp}.

The ansatz many-body state that we introduce is based on the ``exact" many-body state for $L = 0$. The many-body state 
of each component will be a linear superposition of the ``Fock" states of the form 
\begin{eqnarray}
|m_{\rm min}^{N_{m_{\rm min}}^A}, \dots, m_{\rm max}^{N_{m_{\rm max}}^A} \rangle_A \bigotimes 
|m_{\rm min}^{N_{m_{\rm min}}^B}, \dots, m_{\rm max}^{N_{m_{\rm max}}^B} \rangle_B
\nonumber \\
\end{eqnarray} 
for some given truncation to the single-particle states with $m_{\rm min} \le m \le m_{\rm max}$, with the obvious 
constraints in each state $\sum_m N_{m}^i = N_i$, with $i=A,B$ and also with $\sum_{m,i} m N_{m}^i = 0$. Then, one may 
excite the center of mass coordinate using the same amplitudes, thus creating the state 
\begin{eqnarray}
|(m_{\rm min}+m_A)^{N_{m_{\rm min}}^A}, \dots, (m_{\rm max}+m_A)^{N_{m_{\rm max}}^A} \rangle_A \bigotimes
\nonumber \\
|(m_{\rm min}+m_B)^{N_{m_{\rm min}}^B}, \dots, (m_{\rm max}+m_B)^{N_{m_{\rm max}}^B} \rangle_B.
\label{ansatz}
\end{eqnarray}
The resulting state has an angular momentum 
\begin{eqnarray}
 L = N_A m_A + N_B m_B.
\label{bl2}
\end{eqnarray}
Also, this state has the same interaction energy as the one with $L=0$, since the matrix elements do not 
depend on the angular momentum of the colliding particles. Its total energy is higher than the total energy of
the many-body state with $L=0$, $E(L=0)$, due to its higher kinetic energy, 
\begin{eqnarray}
 E^{'''} &=& V(L=0) + \sum_m (m+m_A)^2 N_m^A + \sum_m (m+m_B)^2 N_m^B 
 \nonumber \\
&=& E(L=0) + N_A m_A^2 + N_B m_B^2 + 2 (m_A L_A + m_B L_B).
\label{bl1} 
\nonumber \\
\end{eqnarray}
Here $V(L=0)$ is the exact, total, interaction energy of the full many-body state with $L=0$, and $L_A$, $L_B$ is the 
angular momentum of the $A$ and $B$ components of the state with $L = 0$. In general, their sum has to vanish, $L_A + 
L_B = 0$, without each of them vanishing separately. Still, the states with the dominant amplitudes are the ones for 
which $L_A = 0$ and $L_B = 0$, separately, because of the condition $g > g_{AB}$, which is roughly the condition for 
phase co-existence. As a result, 
\begin{eqnarray}
 E^{'''} - E(L=0) \approx N_A m_A^2 + N_B m_B^2, 
 \label{nst}
\end{eqnarray} 
which becomes exact for $g_{AB} = 0$. Equation (\ref{nst}) is also exact within the mean-field approximation, since 
the terms with $L_A \neq 0$ and $L_B \neq 0$ appear due to the depletion. On the other hand, whether the resulting 
(mean-field, or many body) state is the yrast state, depends on the parameters. Finally, we also mention that 
Eq.\,(\ref{nst}) reduces to Eq.\,(\ref{dief}) when $m_A = 1$ and $m_B = -1$, as expected.

From Eqs.\,(\ref{bl2}) and (\ref{bl1}) if follows trivially that when $L$ is an integer multiple of $N$, $L = q N$, 
then $m_A = m_B = q$, in which case Bloch's theorem \cite{FB} holds exactly, even in a two-component system 
\cite{2comp}, $E^{'''}-E_0 = N q^2$. In the case of the ``traditional" Bloch theorem (i.e., in the case of one 
component) starting from the $L=0$ state, by exciting the center of mass motion one gets (exactly) only the states 
with an additional angular momentum which is an integer multiple of the total number of particles $N$. 

On the other hand, in the present case of a two-component system, this procedure allows us to give $L$ any desired 
value, at least when the populations $N_A$ and $N_B$ are relatively prime, otherwise the argument will hold for 
values of $L$ which are integer multiples of their greatest common divisor. Still, the generated states are not 
necessarily the yrast states, but rather they are candidate yrast states.

\subsection{Results of numerical diagonalization}

We turn now to the results that we get from the diagonalization of the many-body Hamiltonian. We consider as a 
first example the case $N_A=16, N_B=4, g_{AA} = g_{BB} = g = 0.1, g_{AB}=0.05$, with $m_{\rm min}=-1$, and 
$m_{\rm max} = 2$ and the results are shown in the Appendix. For $0 < L \le 4 (= N_B)$ we see that indeed the 
angular momentum of the majority component, $A$, is less than 10\% of the total, which is consistent with the 
results of Sec.\,III B. Partly this relatively large value is due to finite-$N$ corrections; increasing $N$ will 
make this number even smaller. For $5 \le L \le 9$ the dominant state of the $B$ (minority) component is $\phi_1$, 
carrying 4 units of angular momentum, while the additional angular momentum is carried by the $A$ component. This 
is because exciting the $B$ component costs kinetic energy. The state with $L=10 \, (=N/2)$ is analysed in detail
below, for $N=10$ and $L = 5 \, (=N/2)$. The rest of the spectrum follows from Bloch's theorem. 

In order to see the effects that we investigate in the present study we turn now to higher couplings using the 
above as a ``reference" example. To achieve a decent convergence we expand the space of single-particle states to 
$m_{\rm min}=-2$, and $m_{\rm max} = 3$, which forces us to reduce the atom number, as otherwise the dimensionality 
of the Hamiltonian matrix explodes. We thus consider $N_A=8, N_B=2, g_{AA} = g_{BB} = g = 1.5, g_{AB}=0.15$. Another 
example, where $g$ and $g_{AB}$ are closer to each other, follows below. 

For $L=0$ and in the space with $m_{\rm min}= -1$ and $m_{\rm max} = 1$ the dimensionality of the 
Hamiltonian matrix is 26, while the lowest eigenenergy is $\approx 38.5864$. For $m_{\rm min}=-2$, and $m_{\rm max} 
= 2$ the dimensionality becomes 457 and the lowest eigenenergy reduces to $\approx 33.8139$, i.e., there is a reduction 
of roughly 14\%. For $m_{\rm min}=-2$, and $m_{\rm max} = 3$ the dimensionality becomes 1163 and the lowest eigenenergy
reduces further to $\approx 32.8452$, i.e., there is a further reduction of roughly 3\%, indicating that although 
convergence has not been achieved, the results are relatively accurate. 

A generic feature of the above problem is that there is a very rapid increase of the dimensionality of the Hilbert 
space as more single-particle states are included, as seen also in the numbers mentioned above. We should also mention 
that in order to satisfy Bloch's theorem for e.g., $0 \le L \le N$ the single-particle states have to be ``symmetric" 
around 1/2. This is the reason why we choose to work, for example, with $m_{\rm min}=-2$, and $m_{\rm max} = 3$. The
fact that we have to increase the single-particle states in pairs makes it even more difficult to investigate the
convergence of our results and to increase the Hilbert space. For example, going e.g., from $m_{\rm min}=-2$, and 
$m_{\rm max} = 3$ to $m_{\rm min}=-3$, and $m_{\rm max} = 4$ may result in a very large increase of the dimensionality 
of the Hamiltonian matrix (for some fixed $N_A$ and $N_B$).

The lowest-energy eigenstate with $L=0$ (and with an eigenenergy equal to $\approx 32.8452$), consists of the following 
four Fock states (with the amplitudes with the largest absolute value)
\vskip1pc
\begin{tabular}{ |c|c|c|c|c|c|c|c|c|c|c|c|c| }
\hline
\multicolumn{1}{ |c| }{} & \multicolumn{6}{ |c| }{Comp. $A$} & \multicolumn{6}{ |c| }{Comp. $B$} \\
\hline
Ampl. & $\phi_{-2}$ & $\phi_{-1}$ & $\phi_0$ & $\phi_1$ & $\phi_2$ & $\phi_3$ & $\phi_{-2}$ & $\phi_{-1}$ & $\phi_0$ & $\phi_1$ & $\phi_2$ & $\phi_3 $ \\ 
\hline
-0.2559 & 1 & 0 & 6 & 0 & 1 & 0 & 0 & 0 & 2 & 0 & 0 & 0 \\ 
\hline
-0.2650 & 0 &0 &8& 0& 0& 0 & 0 & 1& 0& 1& 0& 0 \\
\hline
-0.4326 & 0& 1& 6& 1& 0& 0 &0& 0& 2& 0& 0& 0 \\ 
\hline
0.6465 & 0& 0& 8& 0& 0& 0 &0& 0& 2& 0& 0& 0 \\
\hline
\end{tabular}
\vskip1pc
In the above notation, the Fock state with e.g., 8 ``$A$" atoms in the single-particle state $\phi_0$ 
and 2 ``$B$" atoms in the single-particle state $\phi_0$ has an amplitude 0.6465, etc. 

To understand the arguments which follow, it is instructive to get some insight into the structure of the above
many-body state. The Fock state with the largest amplitude has zero kinetic energy and it puts all 8 ``$A$" atoms 
at the $m=0$ state, as well as all 2 ``$B$" atoms at the state with $m=0$, also. The following three have a kinetic 
energy which is equal to $2$, $2$, and $8$, respectively. The degeneracy between the first two is lifted by the
interactions. More specifically, in the two specific states there are processes where atoms are transferred from 
the $m=0$ state to the states with $m=\pm 1$, $m=\pm 2$, etc., which lower the energy (they are off-diagonal matrix 
elements which come from, e.g., ${\hat c}_0^2 {\hat c}_{-1}^{\dagger} {\hat c}_1^{\dagger}$ \cite{KMP}). 
  
For $L = 1$, the lowest eigenenergy is $\approx 34.6431$, while the states with the four largest amplitudes are
\vskip1pc
\begin{tabular}{ |c|c|c|c|c|c|c|c|c|c|c|c|c| }
\hline
\multicolumn{1}{ |c| }{} & \multicolumn{6}{ |c| }{Comp. $A$} & \multicolumn{6}{ |c| }{Comp. $B$} \\
\hline
Ampl. & $\phi_{-2}$ & $\phi_{-1}$ & $\phi_0$ & $\phi_1$ & $\phi_2$ & $\phi_3$ & $\phi_{-2}$ & $\phi_{-1}$ & $\phi_0$ & $\phi_1$ & $\phi_2$ & $\phi_3 $ \\ 
\hline
-0.2506 &1& 0& 6& 0& 1& 0 &0 & 0& 1& 1& 0& 0 \\ 
\hline
-0.2904 &0 &0 & 8& 0& 0& 0 &0 & 1& 0& 0& 1& 0 \\
\hline
-0.4223 & 0& 1& 6& 1& 0& 0& 0& 0& 1& 1& 0& 0 \\ 
\hline
0.6323 & 0& 0& 8& 0& 0& 0& 0& 0& 1& 1& 0& 0 \\
\hline
\end{tabular}
\vskip1pc 
Here we see that indeed it is the minority component that carries the angular momentum (in all four Fock states).

For $L = 2$, the lowest eigenenergy is $\approx 34.8276$, with
\vskip1pc
\begin{tabular}{ |c|c|c|c|c|c|c|c|c|c|c|c|c| }
\hline
\multicolumn{1}{ |c| }{} & \multicolumn{6}{ |c| }{Comp. $A$} & \multicolumn{6}{ |c| }{Comp. $B$} \\
\hline
Ampl. & $\phi_{-2}$ & $\phi_{-1}$ & $\phi_0$ & $\phi_1$ & $\phi_2$ & $\phi_3$ & $\phi_{-2}$ & $\phi_{-1}$ & $\phi_0$ & $\phi_1$ & $\phi_2$ & $\phi_3 $ \\ 
\hline
-0.2557 &1& 0& 6& 0& 1& 0 &0& 0& 0& 2& 0& 0 \\ 
\hline
-0.2644 & 1& 0& 6& 0& 1& 0 & 0& 0& 1& 0& 1& 0 \\
\hline
-0.4322 & 0& 1& 6& 1& 0& 0 & 0& 0& 0& 2& 0& 0 \\ 
\hline
0.6461 & 0& 0& 8& 0& 0& 0& 0& 0& 0& 2& 0& 0\\
\hline
\end{tabular}
\vskip1pc
The minority, $B$, component still carries the angular momentum (in all four Fock states). The state with the 
largest amplitude is the one expected also from the mean-field approximation. Furthermore, this state does indeed
result (to high accuracy) from the one with $L=0$ by exciting the center of mass coordinate of the minority 
component, while the difference between the eigenenergy of this state and the one with $L=0$ is $\approx 1.9814$, 
i.e., very close to the value $2 (= N_B)$. These are in agreement with the results presented in Sec.\,III. 

For $L = 3$ the lowest eigenenergy is $\approx 38.7296$, with
\vskip1pc
\begin{tabular}{ |c|c|c|c|c|c|c|c|c|c|c|c|c| }
\hline
\multicolumn{1}{ |c| }{} & \multicolumn{6}{ |c| }{Comp. $A$} & \multicolumn{6}{ |c| }{Comp. $B$} \\
\hline
Ampl. & $\phi_{-2}$ & $\phi_{-1}$ & $\phi_0$ & $\phi_1$ & $\phi_2$ & $\phi_3$ & $\phi_{-2}$ & $\phi_{-1}$ & $\phi_0$ & $\phi_1$ & $\phi_2$ & $\phi_3 $ \\ 
\hline
-0.2401 & 1& 0& 6& 0& 1& 0& 0& 0& 0& 1& 1& 0 \\ 
\hline
-0.3224 & 0& 0& 8& 0& 0& 0 &0& 0& 1& 0& 0& 1 \\
\hline
-0.4035 & 0& 1& 6& 1& 0& 0 &0& 0& 0& 1& 1& 0 \\ 
\hline
0.6057 & 0& 0& 8& 0& 0& 0 &0& 0& 0& 1& 1& 0\\
\hline
\end{tabular}
\vskip1pc
In agreement with the results of Sec.\,III, and contrary to the corresponding state with $L = 5$ given in the Appendix, 
the above state results to rather high accuracy from the one with $L=1$ by exciting the center of mass of the minority 
component. The energy difference is $\approx 4.0865$, while the one predicted by the results of Sec.\,III~C is $2 L - 
N_B = 4$.

For $L = 4$ the lowest eigenenergy is $\approx 40.8526$, with
\vskip1pc
\begin{tabular}{ |c|c|c|c|c|c|c|c|c|c|c|c|c| }
\hline
\multicolumn{1}{ |c| }{} & \multicolumn{6}{ |c| }{Comp. $A$} & \multicolumn{6}{ |c| }{Comp. $B$} \\
\hline
Ampl. & $\phi_{-2}$ & $\phi_{-1}$ & $\phi_0$ & $\phi_1$ & $\phi_2$ & $\phi_3$ & $\phi_{-2}$ & $\phi_{-1}$ & $\phi_0$ & $\phi_1$ & $\phi_2$ & $\phi_3 $ \\ 
\hline
-0.2494 & 1& 0& 6& 0& 1& 0& 0& 0& 0& 0& 2& 0\\ 
\hline
-0.2970 & 0& 0& 8& 0& 0&  0 & 0& 0& 0& 1& 0& 1 \\
\hline
-0.4218 & 0& 1& 6& 1& 0& 0& 0& 0& 0& 0& 2& 0 \\ 
\hline
0.6305 & 0& 0& 8& 0& 0& 0& 0& 0& 0& 0& 2& 0\\
\hline
\end{tabular}
\vskip1pc
Here we observe that the Fock state with the dominant amplitude is the one where all 8 ``$A$" atoms occupy the 
$m=0$ state, as well as all 2 ``$B$" atoms occupy the state with $m=2$. Again, this state results approximately
from the states with $L=0$ and $L=2$, by exciting the center of mass motion of the minority component. The energy
difference between this state and the one with $L=0$ is $\approx 8.0074$, while the one that one gets from 
Sec.\,III C is 8. We stress that for weaker interactions the many-body state does not have the structure seen
above. For example, the state with $L = 8 (= 2 N_B)$ in the Appendix is not of this form, where the state with
the largest amplitude is $0.6158 \, |0,12,4,0 \rangle_A \, |0,0,4,0 \rangle_B$. 

For $L = 5$ the lowest eigenenergy is $\approx 45.7010$, with
\vskip1pc
\begin{tabular}{ |c|c|c|c|c|c|c|c|c|c|c|c|c| }
\hline
\multicolumn{1}{ |c| }{} & \multicolumn{6}{ |c| }{Comp. $A$} & \multicolumn{6}{ |c| }{Comp. $B$} \\
\hline
Ampl. & $\phi_{-2}$ & $\phi_{-1}$ & $\phi_0$ & $\phi_1$ & $\phi_2$ & $\phi_3$ & $\phi_{-2}$ & $\phi_{-1}$ & $\phi_0$ & $\phi_1$ & $\phi_2$ & $\phi_3 $ \\ 
\hline
0.3194  &0& 1& 5& 2& 0& 0& 0& 0& 0& 0& 2& 0\\ 
\hline
0.3194 & 0& 0& 2& 5& 1& 0& 0& 2& 0& 0& 0& 0 \\
\hline
-0.3516 & 0& 0& 7& 1& 0& 0& 0& 0& 0& 0& 2& 0 \\ 
\hline
-0.3516 & 0& 0& 1& 7& 0& 0& 0& 2& 0& 0& 0& 0\\
\hline
\end{tabular}
\vskip1pc
Interestingly, this state with $L = N/2 = 5$ cannot in any way be linked to any other state and it does not result 
from exciting the center of mass motion \cite{comment}. This is seen by comparing this eigenstate with the ones with 
$L=1$ and $L=3$. The state that one would construct following this rule has an energy equal to $\approx 46.8276$, 
which is higher than the actual eigenenergy. Therefore, the system manages to construct a state that lies lower in 
energy. We should recall here that within the mean-field approximation for $L = N/2$ one gets a ``dark" solitary 
wave in the minority component, and the winding number changes. 

Furthermore, this eigenstate has the peculiar feature that the Fock states go in pairs, having the same amplitudes 
(modulo signs). This can be seen by the fact that for every Fock state, there has to be another one, which is its 
mirror image that results from the transformation $m \rightarrow 1 - m$. The first state will have an angular 
momentum $\sum m N_m = N/2$, while the other one $\sum (1-m) N_m = N - L = N/2$. Furthermore, the kinetic energy 
of the first will be $K = \sum m^2 N_m$, while that of the other will be $\sum (1-m)^2 N_m = K + N - 2 L = K$. 
Since the interaction energy will also be the same, that is the reason that these states go in pairs. 

It is interesting that within the mean-field approximation and for $\ell=1/2$ there are two degenerate 
solutions, with a very different structure in $\phi_A$, i.e., the phase of the order parameter $\Psi_A$ of the 
majority component. For $\ell \to (1/2)^{\pm}$ we get either the one, or the other solution (in practice depending, 
e.g., on the initial condition that we use in the algorithm). This is an example of spontaneous symmetry breaking. 
This symmetry is restored within the method of diagonalization, where, for $\ell = 1/2$, we get a superposition of 
these two states.

Returning to the results from numerical diagonalization, the rest of the spectrum, for $L = 6, \dots, 10$, as well 
for $L > 10$, follows (exactly) from the above states, according to Bloch's theorem, as we have also checked 
numerically. 

\begin{figure}
\includegraphics[width=7.5cm,height=7.cm,angle=0]{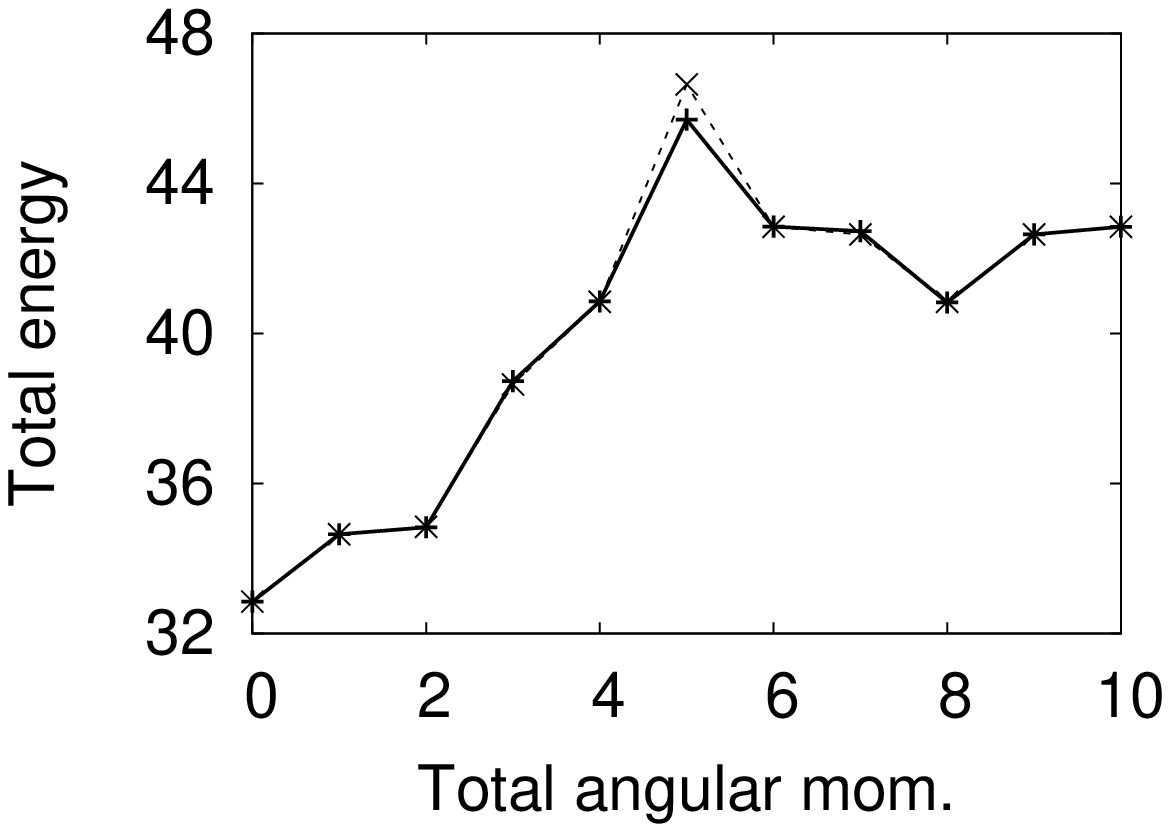}
\includegraphics[width=7.5cm,height=7.cm,angle=0]{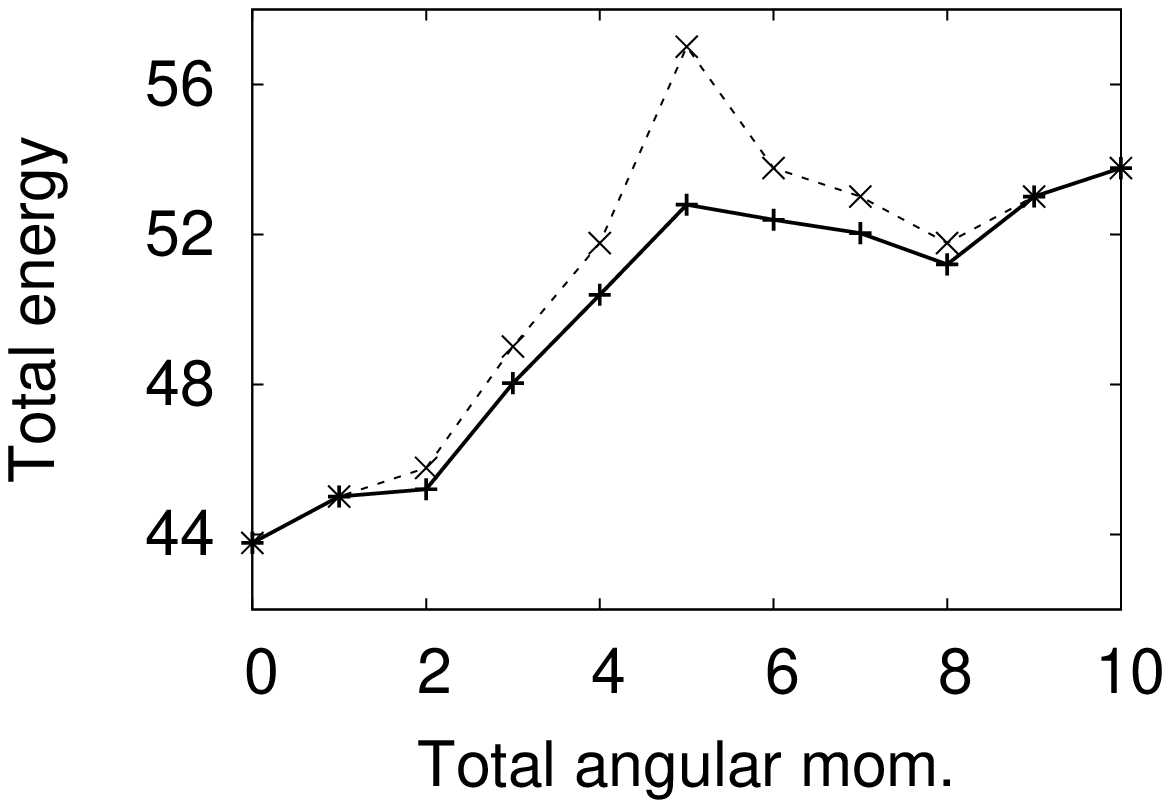}
\vskip3pc
\caption{Top figure: The solid curve connects the lowest eigenenergies, for $N_A=2$, $N_B=8$, $g_{AA}=g_{BB}
=1.5$, and $g_{AB}=0.15$, for $L = 0$ up to 10, in the truncated space $m_{\rm min}=-2$, $m_{\rm max}=3$. 
The dashed curve connects the energies evaluated by the phase transformations described in Sec.\,III C, 
which result from the eigenenergies for $L=0$ and $L=1$. Bottom figure: Same as the top one, with $g_{AB}=0.9$.}
\end{figure}

Another example that we show below has a larger value for $g_{AB}$, $g_{AB}=9/10$, with $N_A=8, N_B=2, g_{AA} = 
g_{BB} = g = 3/2$ being the same as before. The ratio $g/g_{AB}$ is the same as the one in the mean-field calculation 
of Ref.\,\cite{2compp7}. In this study the chosen couplings were rather strong, however here considering the same 
parameters would require inclusion of a large space of single-particle states and a correspondingly huge 
dimensionality of the resulting Hamiltonian matrix. 

The lowest-energy eigenstate with $L=0$ has an eigenenergy equal to $\approx 43.7724$. The Fock states with the four 
largest amplitudes are
\vskip1pc
\begin{tabular}{ |c|c|c|c|c|c|c|c|c|c|c|c|c| }
\hline
\multicolumn{1}{ |c| }{} & \multicolumn{6}{ |c| }{Comp. $A$} & \multicolumn{6}{ |c| }{Comp. $B$} \\
\hline
Ampl. & $\phi_{-2}$ & $\phi_{-1}$ & $\phi_0$ & $\phi_1$ & $\phi_2$ & $\phi_3$ & $\phi_{-2}$ & $\phi_{-1}$ & $\phi_0$ & $\phi_1$ & $\phi_2$ & $\phi_3 $ \\ 
\hline
-0.2420 &1&0&6&0&1&0&0&0&2&0&0&0 \\ 
\hline
0.2422 &0&0&8&0&0&0&0&1&0&1&0&0 \\
\hline
-0.4146 &0&1&6&1&0&0&0&0&2&0&0&0\\ 
\hline
0.6468 & 0&0&8&0&0&0&0&0&2&0&0&0\\
\hline
\end{tabular}
\vskip1pc
For $L = 1$, the lowest eigenenergy is $\approx 45.0110$, while 
\vskip1pc
\begin{tabular}{ |c|c|c|c|c|c|c|c|c|c|c|c|c| }
\hline
\multicolumn{1}{ |c| }{} & \multicolumn{6}{ |c| }{Comp. $A$} & \multicolumn{6}{ |c| }{Comp. $B$} \\
\hline
Ampl. & $\phi_{-2}$ & $\phi_{-1}$ & $\phi_0$ & $\phi_1$ & $\phi_2$ & $\phi_3$ & $\phi_{-2}$ & $\phi_{-1}$ & $\phi_0$ & $\phi_1$ & $\phi_2$ & $\phi_3 $ \\ 
\hline
-0.2312 &0&0&8&0&0&0&0&1&0&0&1&0 \\ 
\hline
-0.2370  &1&0&6&0&1&0&0&0&1&1&0&0 \\
\hline
-0.3676  &0&1&6&1&0&0&0&0&1&1&0&0\\ 
\hline
0.6135   &0&0&8&0&0&0&0&0&1&1&0&0\\
\hline
\end{tabular}
\vskip1pc
Again, we observe that the angular momentum is carried by the minority component. For $L = 2$, the lowest eigenenergy 
is $\approx 45.2024$, with
\vskip1pc
\begin{tabular}{ |c|c|c|c|c|c|c|c|c|c|c|c|c| }
\hline
\multicolumn{1}{ |c| }{} & \multicolumn{6}{ |c| }{Comp. $A$} & \multicolumn{6}{ |c| }{Comp. $B$} \\
\hline
Ampl. & $\phi_{-2}$ & $\phi_{-1}$ & $\phi_0$ & $\phi_1$ & $\phi_2$ & $\phi_3$ & $\phi_{-2}$ & $\phi_{-1}$ & $\phi_0$ & $\phi_1$ & $\phi_2$ & $\phi_3 $ \\ 
\hline
-0.2205 &0&0&8&0&0&0&0&0&1&0&1&0 \\ 
\hline
-0.2369 &1&0&6&0&1&0&0&0&0&2&0&0 \\
\hline
-0.4035 &0&1&6&1&0&0&0&0&0&2&0&0\\ 
\hline
0.6351 &0&0&8&0&0&0&0&0&0&2&0&0\\
\hline
\end{tabular}
\vskip1pc
This state is linked with $|L=0 \rangle$ the way we discussed above. The only difference is that the Fock states with 
the two smallest amplitudes are reversed. For $L = 3$ the lowest eigenenergy is $\approx 48.0354$, with
\vskip1pc
\begin{tabular}{ |c|c|c|c|c|c|c|c|c|c|c|c|c| }
\hline
\multicolumn{1}{ |c| }{} & \multicolumn{6}{ |c| }{Comp. $A$} & \multicolumn{6}{ |c| }{Comp. $B$} \\
\hline
Ampl. & $\phi_{-2}$ & $\phi_{-1}$ & $\phi_0$ & $\phi_1$ & $\phi_2$ & $\phi_3$ & $\phi_{-2}$ & $\phi_{-1}$ & $\phi_0$ & $\phi_1$ & $\phi_2$ & $\phi_3 $ \\ 
\hline
-0.2857 &0&1&6&1&0&0&0&&0&1&1&0 \\ 
\hline
0.3043 &0&1&5&2&0&0&0&0&0&2&0&0 \\
\hline
-0.3555 &0&0&7&1&0&0&0&0&0&2&0&0\\ 
\hline
0.4900  &0&0&8&0&0&0&0&0&0&1&1&0\\
\hline
\end{tabular}
\vskip1pc
The difference between this state and $|L=1 \rangle$ is more pronounced (in the second and the third lines). In 
these two Fock states we observe that there are 2 units of angular momentum, as compared to the first and the 
fourth lines, where there are 3 units of angular momentum, as a result of the increase of $g_{AB}$. Still, the Fock 
state with the largest amplitude is the one expected from the earlier discussion.

For the state with $L = 4$ the lowest eigenenergy is $\approx 50.3904$, with
\vskip1pc
\begin{tabular}{ |c|c|c|c|c|c|c|c|c|c|c|c|c| }
\hline
\multicolumn{1}{ |c| }{} & \multicolumn{6}{ |c| }{Comp. $A$} & \multicolumn{6}{ |c| }{Comp. $B$} \\
\hline
Ampl. & $\phi_{-2}$ & $\phi_{-1}$ & $\phi_0$ & $\phi_1$ & $\phi_2$ & $\phi_3$ & $\phi_{-2}$ & $\phi_{-1}$ & $\phi_0$ & $\phi_1$ & $\phi_2$ & $\phi_3 $ \\ 
\hline
-0.2058 &1&0&6&0&1&0&0&0&0&0&2&0 \\ 
\hline
-0.2074 &0&0&7&1&0&0&0&0&0&1&1&0 \\
\hline
-0.3488 &0&1&6&1&0&0&0&0&0&0&2&0\\ 
\hline
0.5529 &0&0&8&0&0&0&0&0&0&0&2&0\\
\hline
\end{tabular}
\vskip1pc
Again, this state is linked with the states $|L=0 \rangle$ and $|L=2 \rangle$, with the main difference in the third
Fock state, which has 3 units of angular momentum, while the other ones have 4 units. Finally, for $L = 5$ the lowest 
eigenenergy is $\approx 52.7947$, with
\vskip1pc
\begin{tabular}{ |c|c|c|c|c|c|c|c|c|c|c|c|c| }
\hline
\multicolumn{1}{ |c| }{} & \multicolumn{6}{ |c| }{Comp. $A$} & \multicolumn{6}{ |c| }{Comp. $B$} \\
\hline
Ampl. & $\phi_{-2}$ & $\phi_{-1}$ & $\phi_0$ & $\phi_1$ & $\phi_2$ & $\phi_3$ & $\phi_{-2}$ & $\phi_{-1}$ & $\phi_0$ & $\phi_1$ & $\phi_2$ & $\phi_3 $ \\ 
\hline
-0.1817 &0&1&4&3&0&0&0&0&0&1&1&0 \\ 
\hline
-0.1817 &0&0&3&4&1&0&0&1&1&0&0&0\\
\hline
0.1920 &0&0&4&4&0&0&0&0&1&1&0&0\\ 
\hline
0.2038 &0&0&6&2&0&0&0&0&0&1&1&0\\
\hline
0.2038 &0&0&2&6&0&0&0&1&1&0&0&0\\
\hline
\end{tabular}
\vskip1pc
which still is not linked with the other states. 

Figure 5 shows the eigenenergies for $0 \le L \le 10$ for the two values of $g_{AB}$. In the same figure we have 
also used the eigenenergies for $L=0$ and $L=1$ and evaluated the other ones using the arguments presented in 
Sec.\,III C. The agreement for the lower value of $g_{AB}$ is better. With increasing $g_{AB}$ the two systems 
become more coupled and as a result there are processes like, e.g., ${\hat c}_0 {\hat c}_1^{\dagger} 
{\hat d}_0^{\dagger} {\hat d}_1$, which lower the energy and become more important. These processes make the 
amplitudes of the Fock states which constitute the $L=0$ yrast state and have $L_A \neq 0$ and $L_B \neq 0$ 
(with $L_A + L_B =0$) larger. These states are responsible for the observed deviations [see Eq.\,(\ref{bl1})]. 
We also observe the relatively large deviation that appears for $L = 5 = N/2$. This deviation is due to the fact 
that this eigenstate does not result from the other ones via excitation of the center of mass motion.

To conclude, interestingly enough, essentially the whole excitation spectrum (with the exception of the distinct
values of $L = N/2 + N q$, with $q$ being an integer), can thus be derived by the states $L = 0$ and $L = 1$ only 
-- at least approximately -- very much the same way that we saw in Sec.\,III. 

\section{A conjecture: Dispersion relation based on the minimization of the kinetic energy}

As we argued in Sec.\,IV B, starting from the many-body state of a system with $L=0$ it is possible to create
a many-body state with some nonzero value of $L$ at the expense of kinetic energy only, which is of order $N$ 
(in the total energy of the system). Alternatively the many-body state may result from single-particle excitation  
with an energy expense in the interaction energy which is of order $N g$ (still in the total energy of the system), 
for $g_{AA} \approx g_{BB} \approx g_{AB}$, and equal to $g$. Furthermore, for sufficiently strong interactions, 
i.e., when $g$ becomes of order $N$, the system enters the Tonks-Girardeau regime, where the energy does not depend 
on $g$, which is not desirable.

Therefore, provided that 
\begin{eqnarray}
 N \ll N g \ll N^2,
\label{eqq3}
\end{eqnarray}
it may be energetically favorable for the system to carry its angular momentum via the collective excitation described
above. In this case, provided that $N_A$ and $N_B$ are relatively prime one may achieve any value of $L = m N_A + n N_B$. 
The integers $(m, n)$ are the ones which minimize the kinetic energy per particle
\begin{eqnarray}
  K = m^2 N_A + n^2 N_B,
\label{ke}
\end{eqnarray}
under the obvious constraint
\begin{eqnarray}
  L = m N_A + n N_B.
\label{angm}
\end{eqnarray}
Self-consistency requires that the resulting integers $m$ and $n$ have to be of order unity. 

\begin{figure}
\includegraphics[width=7cm,height=5.cm]{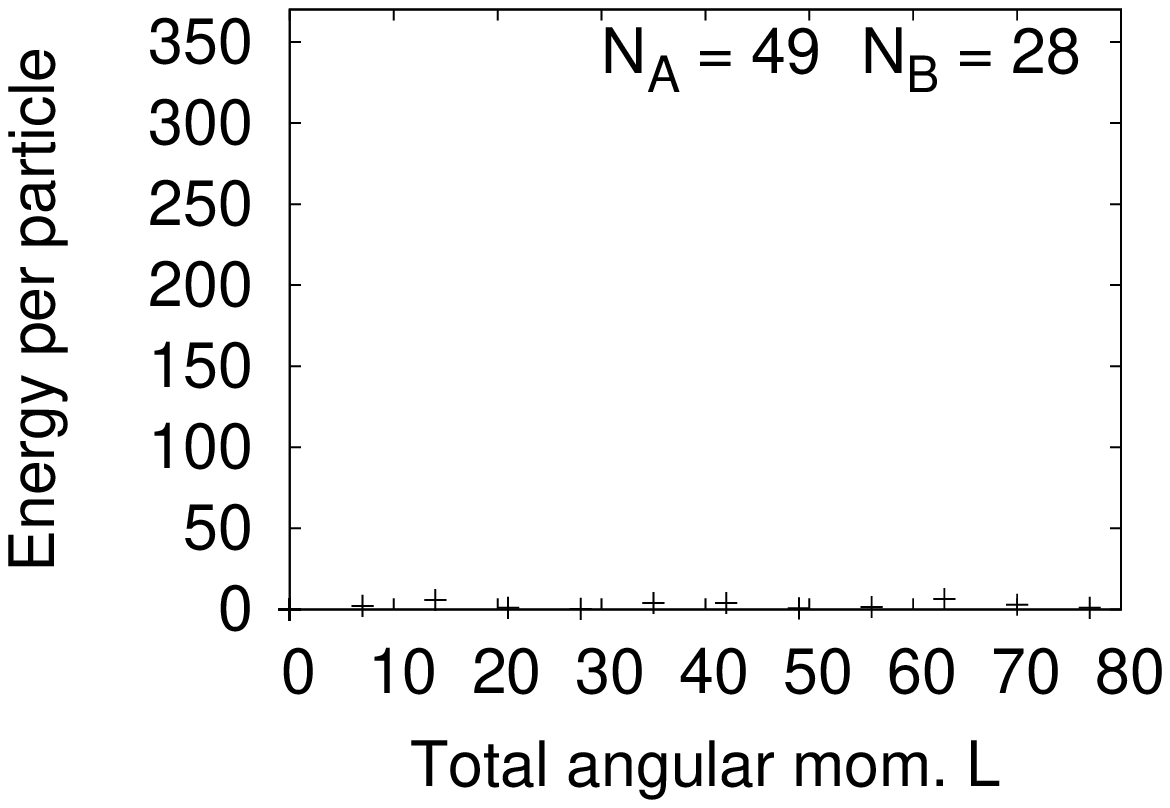}
\includegraphics[width=7cm,height=5.cm]{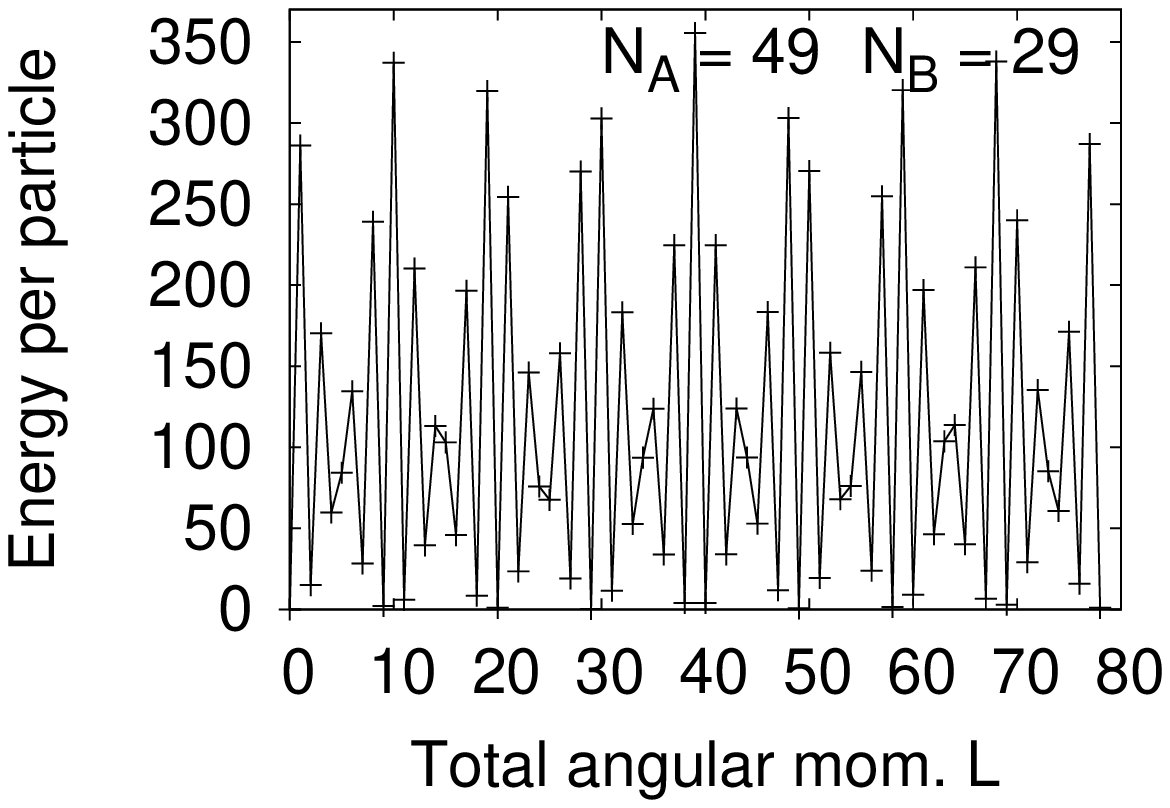}
\includegraphics[width=7cm,height=5.cm]{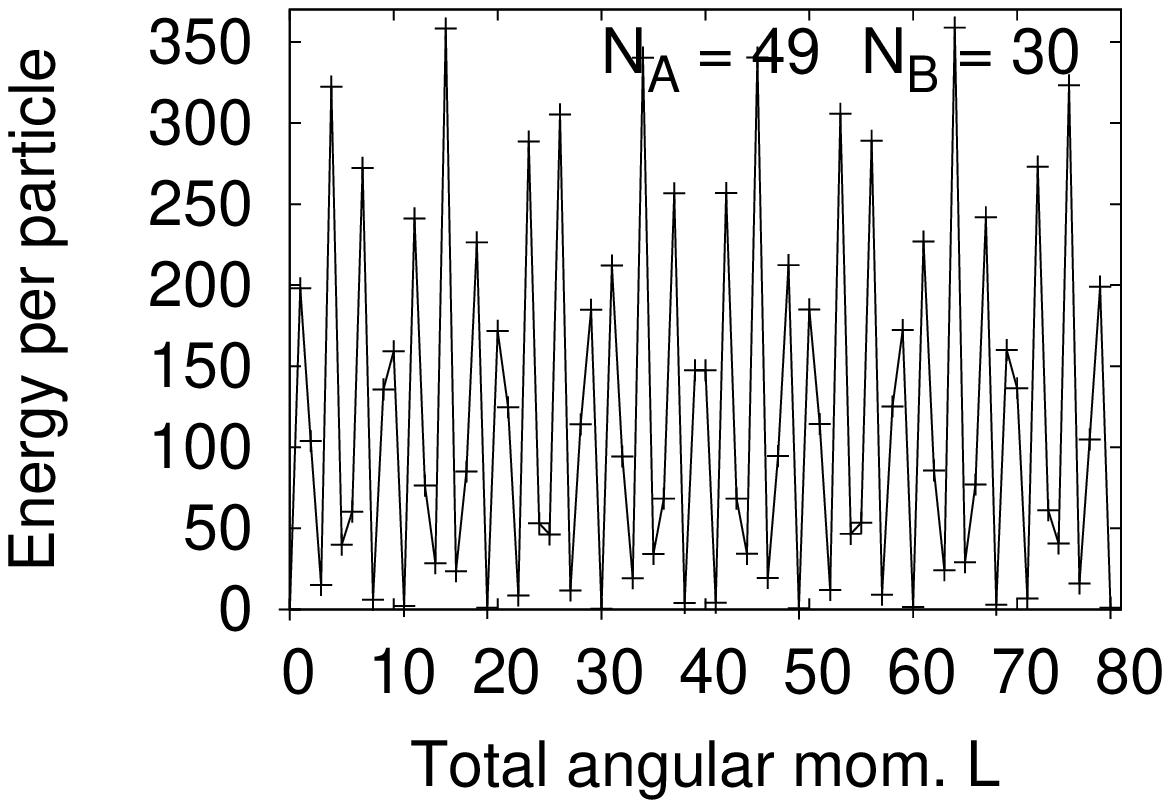}
\includegraphics[width=7cm,height=5.cm]{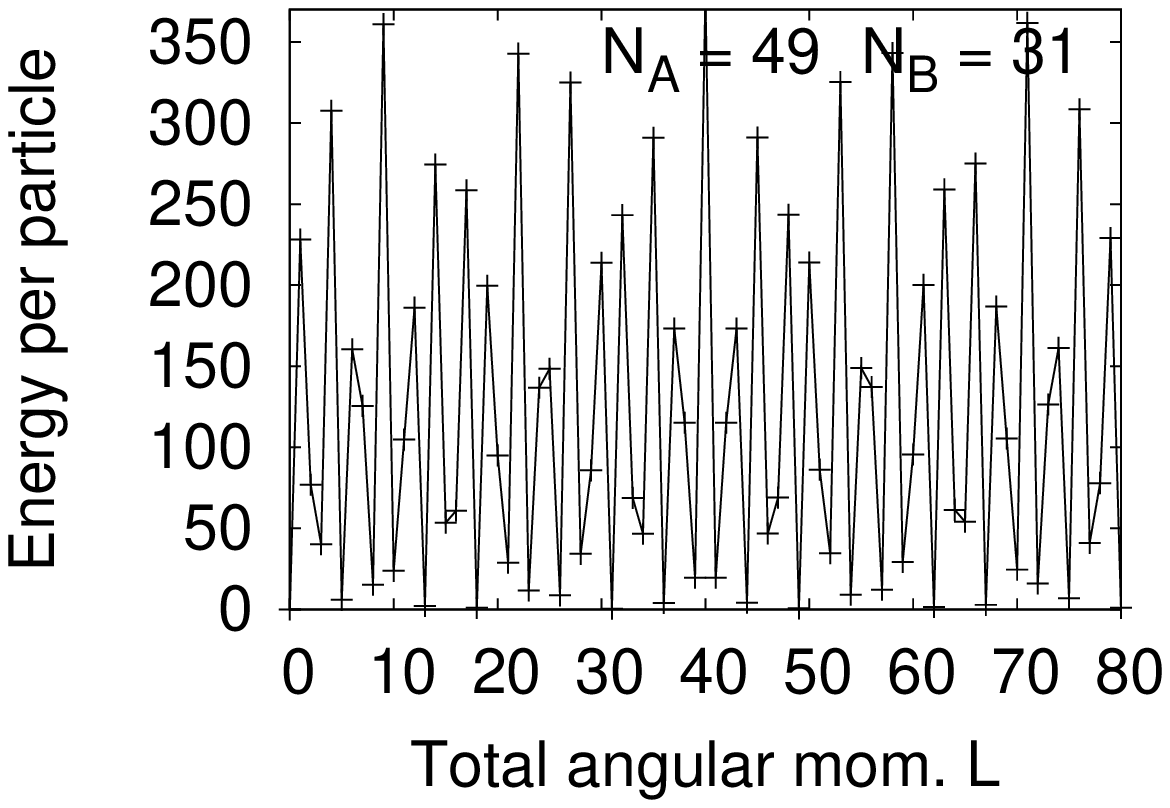}
\vskip2pc
\caption{The dispersion relation (i.e., the kinetic energy) evaluated from the minimization of Eq.\,(\ref{ke}) 
under the constraint of Eq.\,(\ref{angm}), for the numbers of $N_A$ and $N_B$ shown in each plot.}
\end{figure}

\begin{figure}
\includegraphics[width=7cm,height=5.cm]{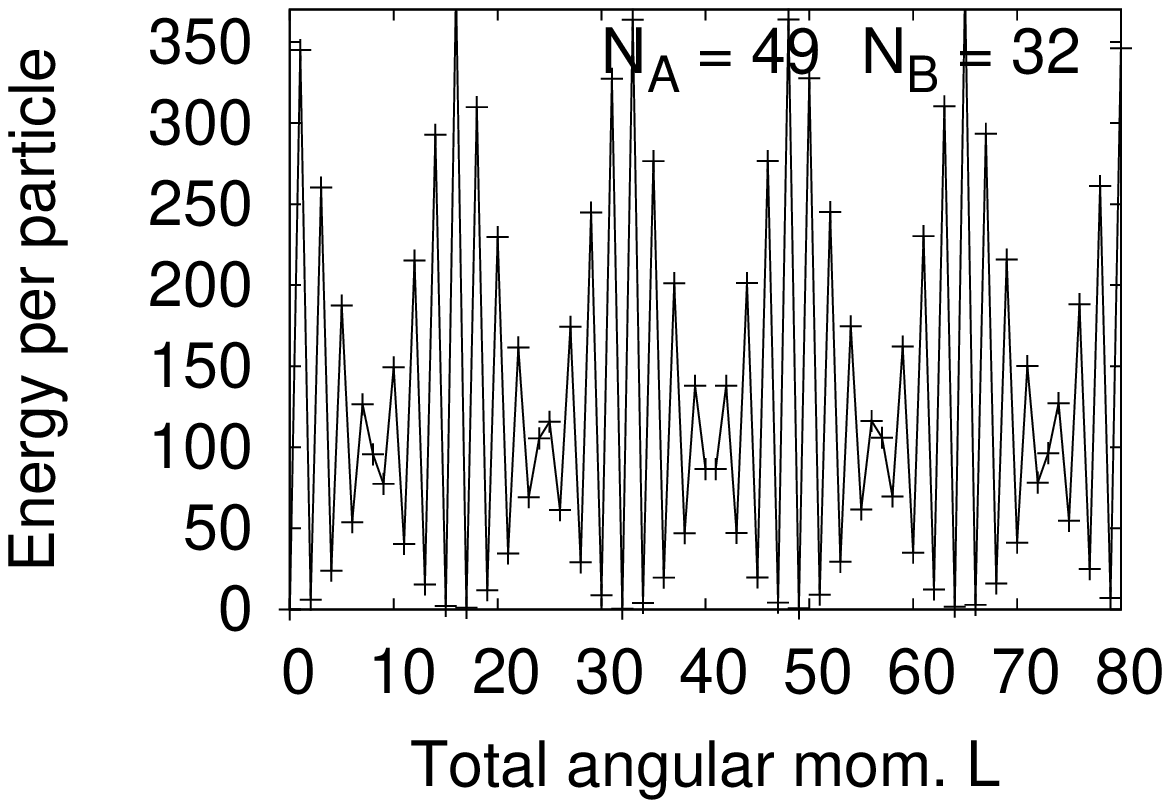}
\includegraphics[width=7cm,height=5.cm]{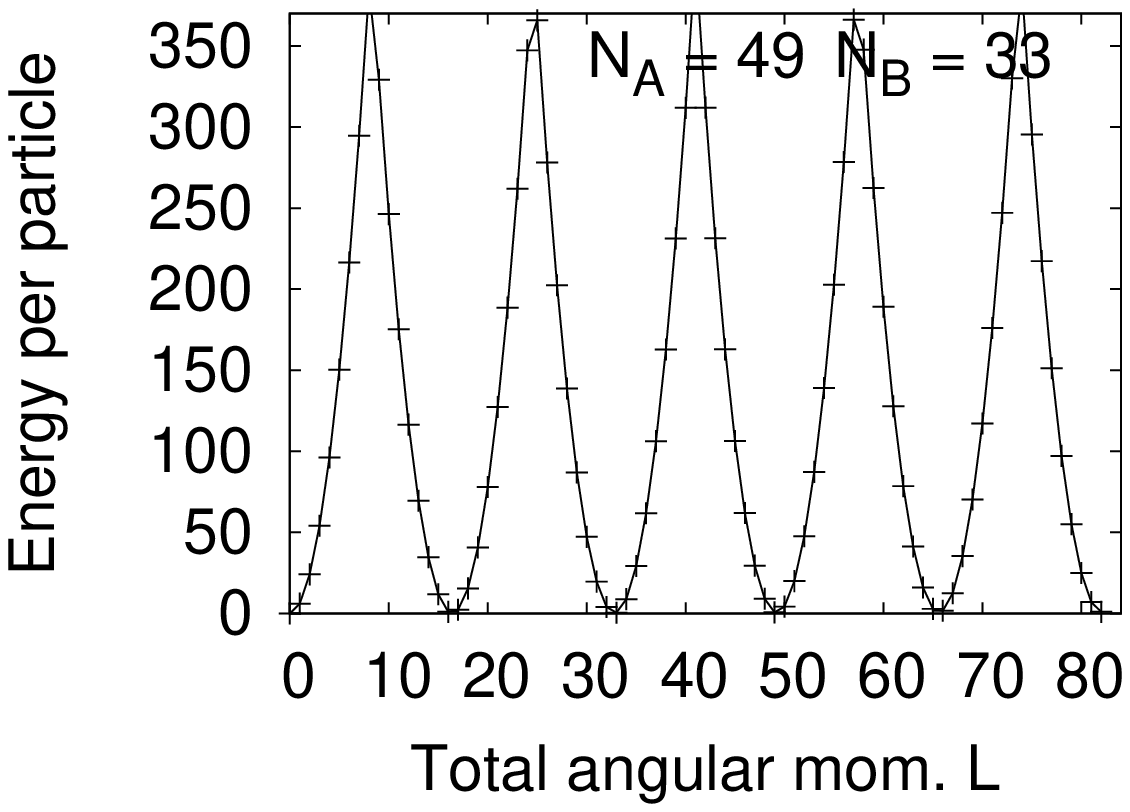}
\includegraphics[width=7cm,height=5.cm]{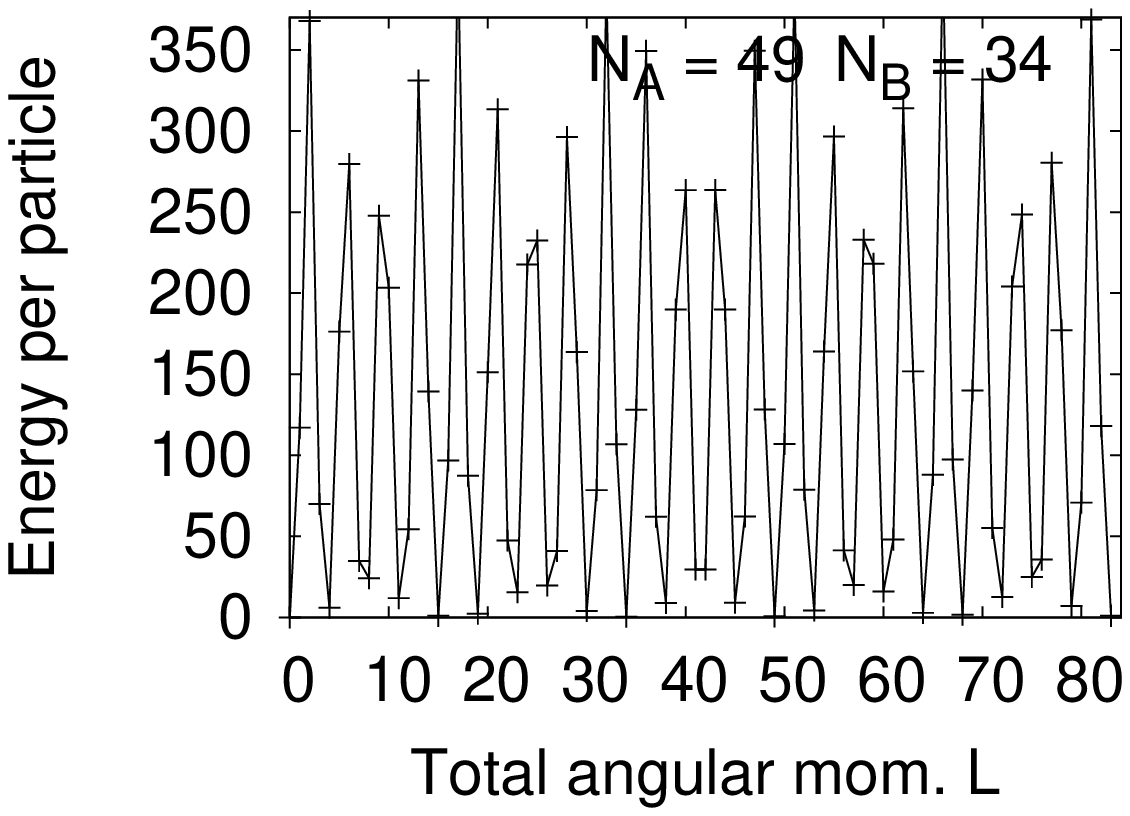}
\includegraphics[width=7cm,height=5.cm]{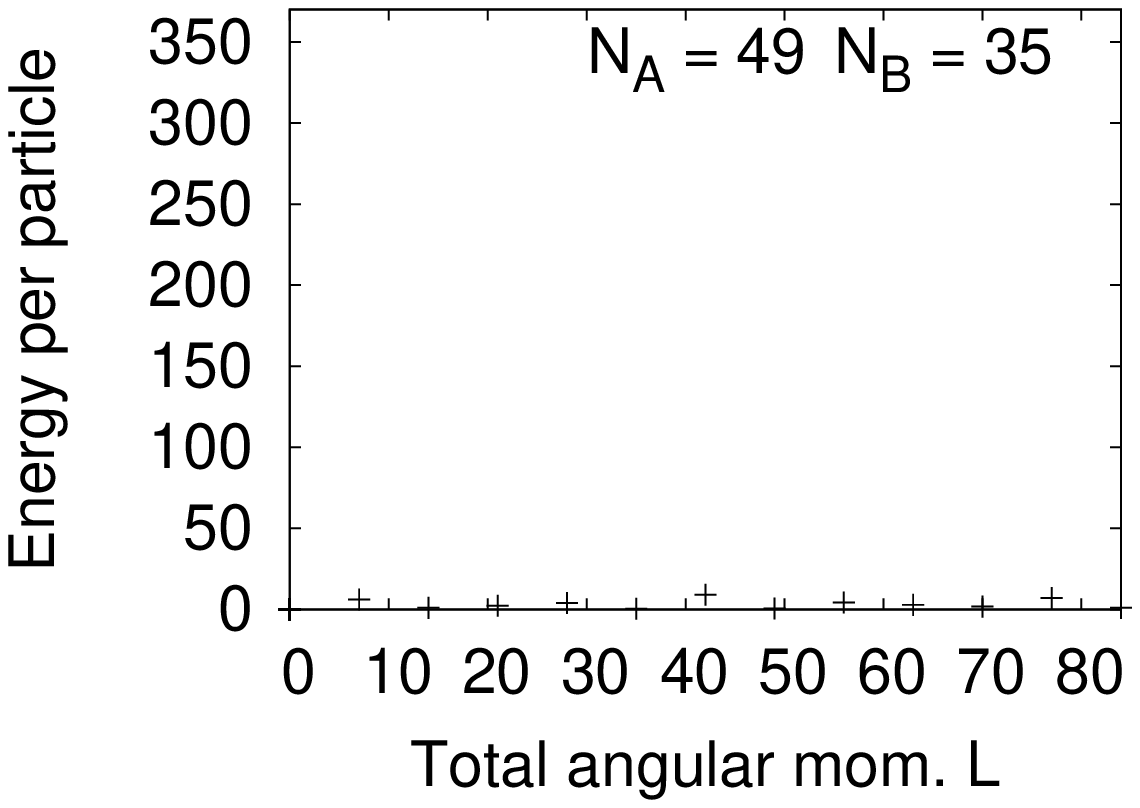}
\vskip2pc
\caption{Same as in Fig.\,6.}
\end{figure}

It is important to point out that Eq.\,(\ref{ke}) and (\ref{angm}) are linear in $N_A$ and $N_B$. Thus, scaling 
$N_A$ and $N_B$ the same way will leave the resulting integers $m$ and $n$ unaffected. On the other hand, 
Eq.\,(\ref{eqq3}) will always be satisfied for a sufficiently large value of $N = N_A + N_B$, for some fixed~$g$. 

The inequality of Eq.\,(\ref{eqq3}) implies that in order for each term to differ by, e.g., one order of magnitude, 
$g$ has to be at least 10, while $N$ has to be at least 100. This introduces a very serious problem in the method 
of numerical diagonalization that we use. Convergence of the results requires that the space that one should work 
with is $|m_{\rm min}| \approx m_{\rm max} \approx \sqrt {N g} \approx 30$. This implies that the dimensionality 
of the resulting matrices is too large and certainly beyond the capability of current technology. 
 
Still, if one could reach these parameters -- which is certainly possible experimentally -- there is an interesting 
behaviour, which we investigate below. The most interesting aspect is that under the conditions presented above, the 
yrast spectrum is determined from the minimization of the kinetic energy and thus becomes trivial. In addition to the 
simplicity of the spectrum, even more interesting is that the dispersion relation may become very sensitive to $N_A$ 
and $N_B$, due to number-theoretic reasons. 

In Figs.\,6 and 7, instead of diagonalizing the many-body Hamiltonian, we minimize Eq.\,(\ref{ke}) under the 
constraint of Eq.\,(\ref{angm}) and plot the dispersion relation [measuring the energy from $E(L=0)$]. As an example, 
we have chosen $N_A = 49$ and $N_B$ from 28 up to 35. For $N_B = 28$, the greatest common divisor of $N_A$ and $N_B$ 
is 7 and for this reason we find a solution for the values of $L$ which are integer multiples of 7, i.e., 0, 7, 14, 
21, 28, 35, 42, 49, 56, 63, 70, and 77. 

For all other values of $L$ the energy will be much higher, so the predicted dispersion relation will have minima 
at these values of $L$. Increasing the population of $N_B$ by one unit, i.e., for $N_B = 29$, the greatest common 
divisor of $N_A$ and $N_B$ is 1. This has dramatic consequences on the dispersion relation, since it is now possible 
to find a solution for all values of $L$ between 0 and $N_A + N_B = 78$. Various interesting patterns show up as $N_B$ 
continues to increase by one unit, until $N_B$ increases by seven units, $N_B = 35$, in which case the greatest 
common divisor of $N_A$ and $N_B$ is again equal to 7, in which case the dispersion shows a similar structure as 
in the case $N_B = 28$. 

A remarkable observation that follows from these results is that even if the population changes by one particle, 
this may change the dispersion dramatically. This is a direct consequence of the number-theoretic nature of the 
problem, much like shell-effects for fermions, due to the Pauli exclusion principle. 

\section{Summary and overview}

In the present study we have studied the dispersion relation of a two-component Bose-Einstein condensed gas that
is confined in a ring potential. The structure of the derived excitation spectrum and the corresponding 
states have immediate consequences on the rotational properties of the system that we have examined and thus they have 
a very interesting physical interpretation. 

To name just the most important ones, we need to recall that the local minima of the dispersion relation correspond 
to non-decaying states, i.e., persistent currents. Furthermore, the states that we have evaluated correspond to 
``vector" solitary-wave solutions, i.e., density disturbances in both components (see Figs.\,2 and 3) which propagate 
together around the ring without change in their shape. In addition, the slope of the dispersion relation gives the 
velocity of propagation of these waves. Finally, the dispersion relation may be used to predict the behaviour of the 
system as it is driven by some external rotation of the trap and also it allows us to extract the hysteretic behaviour. 

Turning to the more specific properties we have derived, we have shown that, quite generally (and not only within 
the mean-field approximation) under certain and rather typical conditions the whole energy spectrum repeats itself 
in a quasi-periodic way. More specifically, if one knows the spectrum in the range of the angular momentum between 
$L=0$ and $L = N_B$, i.e., the population of the minority component, the rest may be derived by exciting the center 
of mass motion of the two components. 

An interesting result that is directly related with the above is the fact that in this range of angular momentum 
the majority of the angular momentum is carried by the minority component, which is a definite experimental 
prediction. Another interesting physical consequence of these results is that, within the mean-field approximation 
-- when the ``dark" soliton appears (in the minority component), the velocity of propagation of the solitary waves 
changes discontinuously. Furthermore, within the many-body scheme the state with this value of the angular momentum 
has some peculiar properties.

One important observation in the problem we have studied is the fact that the matrix elements that determine the 
interaction do not depend on the angular momentum of the colliding particles. As a result, one may start from the
non-rotating many-body state and use these correlations to build many-body states with some nonzero angular momentum.
In the limit of relatively strong interactions these are possible yrast states. The reason is that the energy expense 
that one pays to give the angular momentum is purely kinetic energy and for sufficiently strong interatomic interactions 
this kind of excitation provides an energetically inexpensive way for the system to carry its angular momentum (since 
the correlations are unaffected).

As a result in this limit it is the kinetic energy that has to be minimized, with the interesting consequence that
the energy spectrum is trivial to calculate. Furthermore, much like non-interacting fermions, due to number-theoretic 
reasons the energy spectrum also becomes very sensitive to the population of the two components, as well as the angular 
momentum carried by the system. In a sense, this is an indication of ``quantum chaos", where even infinitesimally small 
changes in the number of atoms (i.e., of order unity) have very significant changes in the dispersion relation, and as 
a result in the rotational properties of the system. While we cannot demonstrate this conjecture numerically because 
of the huge dimensionality of the resulting matrices, there are definite predictions, which may be tested experimentally. 

\acknowledgements

We thank Andy Jackson and Stephanie Reimann for useful discussions.

\appendix*
\section{Yrast states for $N_A = 16$ and $N_B = 4$}

Below we give the result for $N_A=16, N_B=4, g_{AA} = g_{BB} = g = 0.1, g_{AB}=0.05$ in the space with
$-1 \le m \le 2$. 

The lowest-energy eigenstate with $L=0$ has an eigenenergy equal to $\approx 15.1799$. Furthermore, 
the dimensionality of the matrix is 846. The states with the four largest amplitudes are
\vskip1pc
\begin{tabular}{ |c|c|c|c|c|c|c|c|c| }
\hline
\multicolumn{1}{ |c| }{} & \multicolumn{4}{ |c| }{Comp. $A$} & \multicolumn{4}{ |c| }{Comp. $B$} \\
\hline
Ampl. & $\phi_{-1}$ & $\phi_0$ & $\phi_1$ & $\phi_2$ & $\phi_{-1}$ & $\phi_0$ & $\phi_1$ & $\phi_2$ \\ 
\hline
0.0944 &2&12&2&0&0&4&0&0 \\ 
\hline
-0.1139 &0&16&0&0&1&2&1&0 \\
\hline
-0.3122 &1&14&1&0&0&4&0&0\\ 
\hline
0.9301 &0&16&0&0&0&4&0&0\\
\hline
\end{tabular}
\vskip1pc
For $L = 1$, the lowest eigenenergy is $\approx 16.3549$, while 
\vskip1pc
\begin{tabular}{ |c|c|c|c|c|c|c|c|c| }
\hline
\multicolumn{1}{ |c| }{} & \multicolumn{4}{ |c| }{Comp. $A$} & \multicolumn{4}{ |c| }{Comp. $B$} \\
\hline
Ampl. & $\phi_{-1}$ & $\phi_0$ & $\phi_1$ & $\phi_2$ & $\phi_{-1}$ & $\phi_0$ & $\phi_1$ & $\phi_2$ \\ 
\hline
0.1221 &1&13&2&0&0&4&0&0 \\ 
\hline
-0.2695 &0&15&1&0&0&4&0&0 \\
\hline
-0.2811 &1&14&1&0&0&3&1&0\\ 
\hline
0.8893 &0&16&0&0&0&3&1&0\\
\hline
\end{tabular}
\vskip1pc
For $L = 2$, the lowest eigenenergy is $\approx 17.4151$, with
\vskip1pc
\begin{tabular}{ |c|c|c|c|c|c|c|c|c| }
\hline
\multicolumn{1}{ |c| }{} & \multicolumn{4}{ |c| }{Comp. $A$} & \multicolumn{4}{ |c| }{Comp. $B$} \\
\hline
Ampl. & $\phi_{-1}$ & $\phi_0$ & $\phi_1$ & $\phi_2$ & $\phi_{-1}$ & $\phi_0$ & $\phi_1$ & $\phi_2$ \\ 
\hline
0.1245 &1&13&2&0&0&3&1&0  \\ 
\hline
-0.2742 &1&14&1&0&0&2&2&0 \\
\hline
-0.2814 &0&15&1&0&0&3&1&0\\ 
\hline
0.8840 &0&16&0&0&0&2&2&0 \\
\hline
\end{tabular}
\vskip1pc
For $L = 3$ the lowest eigenenergy is $\approx 18.3432$, with
\vskip1pc
\begin{tabular}{ |c|c|c|c|c|c|c|c|c| }
\hline
\multicolumn{1}{ |c| }{} & \multicolumn{4}{ |c| }{Comp. $A$} & \multicolumn{4}{ |c| }{Comp. $B$} \\
\hline
Ampl. & $\phi_{-1}$ & $\phi_0$ & $\phi_1$ & $\phi_2$ & $\phi_{-1}$ & $\phi_0$ & $\phi_1$ & $\phi_2$ \\ 
\hline
0.1086 &1&13&2&0&0&2&2&0  \\ 
\hline
-0.2461 &0&15&1&0&0&2&2&0 \\
\hline
-0.2826 &1&14&1&0&0&1&3&0\\ 
\hline
0.8960 &0&16&0&0&0&1&3&0 \\
\hline
\end{tabular}
\vskip1pc
For $L = 4$ the lowest eigenenergy is $\approx 19.1274$, with
\vskip1pc
\begin{tabular}{ |c|c|c|c|c|c|c|c|c| }
\hline
\multicolumn{1}{ |c| }{} & \multicolumn{4}{ |c| }{Comp. $A$} & \multicolumn{4}{ |c| }{Comp. $B$} \\
\hline
Ampl. & $\phi_{-1}$ & $\phi_0$ & $\phi_1$ & $\phi_2$ & $\phi_{-1}$ & $\phi_0$ & $\phi_1$ & $\phi_2$ \\ 
\hline
-0.1025 &0&16&0&0&0&1&2&1  \\ 
\hline
-0.1791 &0&15&1&0&0&1&3&0\\
\hline
-0.3017 &1&14&1&0&0&0&4&0\\ 
\hline
0.9153 &0&16&0&0&0&0&4&0 \\
\hline
\end{tabular}
\vskip1pc
For $L = 5$ the lowest eigenenergy is $\approx 21.0242$, with
\vskip1pc
\begin{tabular}{ |c|c|c|c|c|c|c|c|c| }
\hline
\multicolumn{1}{ |c| }{} & \multicolumn{4}{ |c| }{Comp. $A$} & \multicolumn{4}{ |c| }{Comp. $B$} \\
\hline
Ampl. & $\phi_{-1}$ & $\phi_0$ & $\phi_1$ & $\phi_2$ & $\phi_{-1}$ & $\phi_0$ & $\phi_1$ & $\phi_2$ \\ 
\hline
-0.1791 &0&16&0&0&0&0&3&1  \\ 
\hline
-0.2520 &0&14&2&0&0&1&3&0\\
\hline
-0.3693 &1&13&2&0&0&0&4&0\\ 
\hline
0.8354 &0&15&1&0&0&0&4&0 \\
\hline
\end{tabular}
\vskip1pc
For $L = 6$ the lowest eigenenergy is $\approx 22.8034$, with
\vskip1pc
\begin{tabular}{ |c|c|c|c|c|c|c|c|c| }
\hline
\multicolumn{1}{ |c| }{} & \multicolumn{4}{ |c| }{Comp. $A$} & \multicolumn{4}{ |c| }{Comp. $B$} \\
\hline
Ampl. & $\phi_{-1}$ & $\phi_0$ & $\phi_1$ & $\phi_2$ & $\phi_{-1}$ & $\phi_0$ & $\phi_1$ & $\phi_2$ \\ 
\hline
-0.2058 &0&15&1&0&0&0&3&1  \\ 
\hline
-0.3104 &0&13&3&0&0&1&3&0\\
\hline
-0.3910 &1&12&3&0&0&0&4&0\\ 
\hline
0.7640 &0&14&2&0&0&0&4&0 \\
\hline
\end{tabular}
\vskip1pc
For $L = 7$ the lowest eigenenergy is $\approx 24.4524$, with
\vskip1pc
\begin{tabular}{ |c|c|c|c|c|c|c|c|c| }
\hline
\multicolumn{1}{ |c| }{} & \multicolumn{4}{ |c| }{Comp. $A$} & \multicolumn{4}{ |c| }{Comp. $B$} \\
\hline
Ampl. & $\phi_{-1}$ & $\phi_0$ & $\phi_1$ & $\phi_2$ & $\phi_{-1}$ & $\phi_0$ & $\phi_1$ & $\phi_2$ \\ 
\hline
0.2136 &1&10&5&0&0&1&3&0 \\ 
\hline
-0.3627 &0&12&4&0&0&1&3&0\\
\hline
-0.3855 &1&11&4&0&0&0&4&0\\ 
\hline
0.6938 &0&13&3&0&0&0&4&0 \\
\hline
\end{tabular}
\vskip1pc
For $L = 8$ the lowest eigenenergy is $\approx 25.9611$, with
\vskip1pc
\begin{tabular}{ |c|c|c|c|c|c|c|c|c| }
\hline
\multicolumn{1}{ |c| }{} & \multicolumn{4}{ |c| }{Comp. $A$} & \multicolumn{4}{ |c| }{Comp. $B$} \\
\hline
Ampl. & $\phi_{-1}$ & $\phi_0$ & $\phi_1$ & $\phi_2$ & $\phi_{-1}$ & $\phi_0$ & $\phi_1$ & $\phi_2$ \\ 
\hline
0.2451 &1&9&6&0&0&1&3&0 \\ 
\hline
-0.3568 &1&10&5&0&0&0&4&0\\
\hline
-0.4100 &0&11&5&0&0&1&3&0\\ 
\hline
0.6158 &0&12&4&0&0&0&4&0 \\
\hline
\end{tabular}
\vskip1pc
For $L = 9$ the lowest eigenenergy is $\approx 27.3137$, with
\vskip1pc
\begin{tabular}{ |c|c|c|c|c|c|c|c|c| }
\hline
\multicolumn{1}{ |c| }{} & \multicolumn{4}{ |c| }{Comp. $A$} & \multicolumn{4}{ |c| }{Comp. $B$} \\
\hline
Ampl. & $\phi_{-1}$ & $\phi_0$ & $\phi_1$ & $\phi_2$ & $\phi_{-1}$ & $\phi_0$ & $\phi_1$ & $\phi_2$ \\ 
\hline
0.2956 &1&9&6&0&0&0&4&0 \\ 
\hline
0.2957 &0&9&7&0&0&2&2&0\\
\hline
-0.4401 &0&10&6&0&0&1&3&0\\ 
\hline
0.5067 &0&11&5&0&0&0&4&0 \\
\hline
\end{tabular}
\vskip1pc
Finally, for $L = 10$ the lowest eigenenergy is $\approx 28.4570$, with
\vskip1pc
\begin{tabular}{ |c|c|c|c|c|c|c|c|c| }
\hline
\multicolumn{1}{ |c| }{} & \multicolumn{4}{ |c| }{Comp. $A$} & \multicolumn{4}{ |c| }{Comp. $B$} \\
\hline
Ampl. & $\phi_{-1}$ & $\phi_0$ & $\phi_1$ & $\phi_2$ & $\phi_{-1}$ & $\phi_0$ & $\phi_1$ & $\phi_2$ \\ 
\hline
-0.2882 &0&6&10&0&0&4&0&0 \\ 
\hline
-0.2882 &0&10&6&0&0&0&4&0\\
\hline
0.3625 &0&7&9&0&0&3&1&0\\ 
\hline
0.3625 &0&9&7&0&0&1&3&0 \\
\hline
-0.3784 &0&8&8&0&0&2&2&0 \\
\hline
\end{tabular}


\begin{thebibliography}{99}
\bibitem{comment} A state that resembles this one has been seen in Ref.\,\cite{2compp0}, for $g = g_{AB}$, 
and for weak interactions.
\end{thebibliography}
\end{document}